\documentclass[aps,prd,onecolumn,groupedaddress,nofootinbib,amssymb]{revtex4}
\usepackage[dvips]{graphicx}
\usepackage{amssymb}
\usepackage{amsmath}
\usepackage{graphicx,,color}
\usepackage{amsfonts}
\usepackage{bm}
\usepackage{cancel}
\usepackage{comment}
\usepackage[dvipsnames]{xcolor}
\usepackage{makecell}

\newcommand{\bea}{\begin{eqnarray}}
\newcommand{\eea}{  \end{eqnarray}}

\allowdisplaybreaks[4]

\begin{document}

\tolerance=5000

\title{Analytic Singular Slow-roll Inflation}
\author{V.K. Oikonomou$^{1,2}$}\email{voikonomou@gapps.auth.gr;v.k.oikonomou1979@gmail.com}
\affiliation{$^{1)}$Department of Physics, Aristotle University of
Thessaloniki, Thessaloniki 54124, Greece\\
$^{2)}$ Center for Theoretical Physics, Khazar University, 41
Mehseti Str., Baku, AZ-1096, Azerbaijan}

\begin{abstract}
We study a class of minimally coupled scalar field theories which
leads to analytic solutions for the Hubble rate and the scalar
field, where the scalar field obeys a generalized tracking law
$\dot{\phi}^2\sim H^{-m}$. The inflationary phenomenology for this
class of models can be studied fully analytically. The resulting
phenomenology is compatible with the ACT data and for limiting
cases, the spectral index is bluer than the ACT constraints and
tends to the value $n_{\mathcal{S}}=0.98$, while in the limiting
case, the tensor-to-scalar ratio takes very small values, nearly
zero. In addition, we prove analytically that the phenomenology is
a one-parameter model, and the inflationary observables encode the
scaling exponent $m$ of the generalized kinetic attractor
$\dot{\phi}^2\sim H^{-m}$. Furthermore, the tensor-to-scalar ratio
and the spectral index have a simple linear and $m$-dependent
relation. More importantly, the resulting cosmology describes a
Universe that has a finite scale factor at $t=0$, thus
non-singular, evolves and expands realizing a slow-roll
inflationary era and after that it reaches classically a pressure
singularity. Classically, the Universe can pass through this
singularity, and a turnaround cosmology is realized with the
Universe contracting after the turnaround point. However, before
the singularity is realized classically, the quantum phenomena
dominate the evolution, avoiding the singularity. Specifically we
consider the Nojiri-Odintsov conformal anomaly mechanism and we
qualitatively show that the conformal anomaly erases the classical
singular evolution and at the same time it enhances particle
creation, which eventually reheats the Universe. Thus in this
model the scalar field oscillations and the numerous couplings of
the inflaton to the Standard Model particles are not required for
reheating. In this context, scalar perturbations are enhanced and
thus the formation of primordial black holes and the generation of
secondary gravitational waves is enhanced, which we briefly
discuss qualitatively. We also discuss several other mechanisms
that may lead to the avoidance of the pressure singularity.
\end{abstract}

\maketitle

\section{Introduction}

Inflation \cite{inflation1,inflation2,inflation3,inflation4} is
today the primary theory for describing the primordial era of our
Universe. Theoretically it solves all the shortcomings of the hot
Big Bang theory, such as the horizon problem, the flatness problem
and the monopole problem. To date, inflation is only constrained,
not yet observed, but the existence of a red tilted scalar
spectral index, the existence of a spectrum of nearly Gaussian
perturbations and the confirmed flatness of the Universe already
affirms that the Universe has a quantum origin. The inflationary
era is a classical era which serves as a link between the quantum
gravity dominated Universe and the classical evolutionary regime.
This is one of the reasons that inflation is so interesting from a
theoretical point of view, since quantum gravity phenomena may
leave their imprint on the observables of the inflationary era. In
the next decade, many experiments are expected to further
constrain or even detect the inflationary era. The smoking gun
detection of inflation would be the detection of the B-mode (curl
mode) in the cosmic microwave background (CMB) radiation. The
Simons observatory \cite{SimonsObservatory:2019qwx}, and the
LiteBird \cite{LiteBIRD:2022cnt} and in addition the CMB stage 4
experiments are expected to intensively seek for the B-mode, or
further constrain the inflationary era. In addition to these
experiments, the future gravitational wave experiments
\cite{Hild:2010id,Baker:2019nia,Smith:2019wny,Crowder:2005nr,Smith:2016jqs,Seto:2001qf,Kawamura:2020pcg,Bull:2018lat,LISACosmologyWorkingGroup:2022jok}
will be able to probe the stochastic gravitational wave background
in the Universe, which can also be generated by inflation. In 2023
the pulsar-timing-arrays experiments like NANOGrav already
confirmed the existence of a stochastic gravitational wave
background \cite{NANOGrav:2023gor}, but the detected signal cannot
be explained by inflation
\cite{Vagnozzi:2023lwo,Oikonomou:2023qfz}. In 2025, the ACT data
release stirred things up in inflationary cosmology because the
reported spectral index is in tension with the Planck 2018 data
\cite{Planck:2018jri}. Specifically, the ACT reported spectral
index is constrained as follows \cite{ACT:2025fju, ACT:2025tim},
\begin{align}
\label{act} n_{\mathcal{S}}=0.9743 \pm 0.0034\, .
\end{align}
This must be combined with the updated Planck/BICEP constraints on
the tensor-to-scalar ratio \cite{BICEP:2021xfz},
\begin{align}
\label{planck} r<0.036\, .
\end{align}
The ACT data release had generated a large stream of articles
aiming to reconcile inflationary models with the ACT data, see for
example Refs.
\cite{Kallosh:2025rni,Gao:2025onc,Liu:2025qca,Yogesh:2025wak,Yi:2025dms,Peng:2025bws,Yin:2025rrs,Byrnes:2025kit,
Wolf:2025ecy,Aoki:2025wld,Gao:2025viy,Zahoor:2025nuq,Ferreira:2025lrd,Mohammadi:2025gbu,Choudhury:2025vso,
Odintsov:2025wai,Q:2025ycf,Zhu:2025twm,Kouniatalis:2025orn,Hai:2025wvs,Dioguardi:2025vci,Yuennan:2025kde,
Kuralkar:2025zxr,Kuralkar:2025hoz,Modak:2025bjv,Oikonomou:2025xms,Odintsov:2025jky,
Aoki:2025ywt,Ahghari:2025hfy,McDonough:2025lzo,Chakraborty:2025wqn,NooriGashti:2025gug,Yuennan:2025mlg,
Deb:2025gtk,Afshar:2025ndm,Ellis:2025zrf,Iacconi:2025odq,Yuennan:2025tyx,Wang:2025cpp,Qiu:2025uot,Wang:2025dbj,Asaka:2015vza,Oikonomou:2025htz,Choudhury:2025hnu,Singh:2025uyr,Kim:2025dyi,Peng:2026ofs}.

In this work we aim to present a class of analytic inflationary
models, in the context of minimally coupled scalar field theory,
which naturally produce a bluer spectral index compared to
standard inflationary models. So these models are naturally
compatible with the ACT data, and perfectly fitted within the
constraints. In fact the limiting cases of the models we shall
present tend to increase the tilt of the spectral index around the
value $n_{\mathcal{S}}\sim 0.98$. The class of models we shall
present assume that the kinetic energy of the scalar field obeys
$\dot{\phi}^2\sim H^{-m}$ in a flat Friedmann-Robertson-Walker
spacetime, with $m$ an even positive integer. With this
constraint, the field equations of the minimally coupled scalar
field theories are solved analytically, so we obtain the Hubble
rate and the scalar field as a function of the cosmic time. The
model can be investigated fully analytically, and thus we have
full command on the dynamics of inflation. The resulting
expressions for the spectral index and the tensor-to-scalar ratio
are particularly simple and depend only on one parameter, the
parameter $m$ and of course on the $e$-foldings number. So the
class of analytic models we study is a one-parameter class of
models. The physics of the analytic scalar inflation is quite
interesting, since the Universe starts from time instance $t=0$
with a finite scale factor, thus the Universe in this context does
not start from a Big Bang singularity. As time evolves, a
slow-roll inflationary regime is realized, which as we prove, is
compatible with the ACT data. In fact as $m$ increases, the
spectral index settles in the value $n_{\mathcal{S}}\sim 0.98$ and
the tensor-to-scalar ratio tends to zero. After the inflationary
era ends, the cosmological system develops classically a pressure
singularity, also known as Type II singularity according to the
Nojiri-Odintsov-Tsujikawa classification \cite{Nojiri:2005sx}.
This is not a crushing type singularity, so the Universe
classically can pass through this singularity, and realize a
turnaround cosmology, with the scale factor decreasing
exponentially and the Universe starts to contract. However the
classical description is known to collapse once finite-time
singularities are approached. Thus the quantum phenomena are
initiated before the singularity is reached, thus we have this
physically appealing picture that after the inflationary era ends,
the Universe approaches the finite-time singularity and quantum
phenomena start to dominate. One of the possible scenarios is that
the Nojiri-Odintsov conformal anomaly scenario takes place
\cite{Nojiri:2020sti}. In this case, the conformal anomaly due to
massless particles, contributes to the Einstein field equations
and thus the classically singular solution $\dot{\phi}^2\sim
H^{-m}$ no longer describes the evolution of the Universe. Thus
the singularity is resolved quantum mechanically. The effect of
the quantum phenomena also leads to extreme particle creation,
which eventually reheats the Universe. In addition to these
phenomena, near the singularity, the scalar perturbations are
amplified, and this may lead to the formation of primordial black
holes or even generate secondary gravitational waves with enhanced
energy spectrum. Another perspective for the resolution of the
classical singularity is the effective inflationary field theory
approach, in the context of which, once the background solution
$\dot{\phi}^2$ reaches a cutoff, the classical scalar theory
ceases to describe the evolution of the Universe. We qualitatively
discuss this perspective, and in addition to this, we also briefly
discuss the possibility of having inflationary phase transitions.

In the rest of this article, we shall assume that the background
metric is a flat Friedmann-Robertson-Walker (FRW) metric, with
line element,
\begin{align}
\label{metricflrw} ds^2 = - dt^2 + a(t)^2 \sum_{i=1,2,3}
\left(dx^i\right)^2\, .
\end{align}

\section{The Class of Analytic Slow-roll Inflation and the Cosmological Perturbations}

In this section we shall present the class of analytic solutions
we obtained in minimally coupled scalar field theory. We shall
discuss the constraints required for the consistency of the
solutions and we shall obtain the spectral index of the scalar
perturbations, and the tensor-to-scalar ratio as functions of the
slow-roll parameters associated with scalar inflation, directly
from the cosmological perturbations, pinpointing the critical
assumptions required for the derivation of the final formulas. As
we will show, the resulting relations of the observational indices
are identical to those of minimally coupled scalar field theory.
We shall focus on the class of analytic slow-roll inflation that
yields results compatible with the recent ACT data, but we shall
also discuss alternative scenarios which shall be briefly analyzed
in the end of this section. We shall consider a pure minimally
coupled scalar field theory in the absence of perfect matter
fluids with gravitational action,
\begin{equation}
    \centering\label{act}
    S = \int d^4 x \sqrt{- g}\left(\frac{R}{2 \kappa^2} - \frac{1}{2}g^{\mu \nu} \partial_\mu \phi \partial_\nu \phi - V(\phi) \right),
\end{equation}
and by varying the action with respect to the scalar field and the
metric for a FRW metric, we obtain the scalar field equation,
\begin{equation}
    \centering\label{eomphi}
    \ddot{\phi} + 3 H \dot{\phi} + V' = 0\, .
\end{equation}
the Friedmann equation,
\begin{equation}
    \centering\label{fr1}
    \frac{3}{\kappa^2} H^2 =\frac{1}{2}\dot{\phi}^2 + V(\phi),
\end{equation}
and the Raychaudhuri equation,
\begin{equation}
    \centering\label{fr2}
    \frac{2}{\kappa^2}\dot{H} =-\dot{\phi}^2.
\end{equation}
No assumption is made for the slow-roll conditions. In the next
subsection, we shall provide analytic solutions of the field
equations.

\subsection{Analytic Slow-roll Inflation Realization}

Our approach for obtaining analytic slow-roll solutions for
minimally coupled scalar field inflation is by assuming that the
derivative of the scalar field is a function of the Hubble rate of
the form,
\begin{equation}\label{mainequationpro}
\dot{\phi}^2=\frac{\beta}{\kappa^4}\left(\frac{H(t)}{H_0}\right)^{-m}\,
,
\end{equation}
where $\kappa^2=\frac{1}{M_p^2}$ and $M_p$ is the reduced Planck
mass, $\beta$ is a dimensionless parameter and also $H_0$ is some
arbitrary mass scale with mass dimensions $[H_0]=[m]$. For
convenience we shall introduce the dimensionful parameter
$\gamma$, as follows,
\begin{equation}\label{gammadef}
\gamma=\frac{\beta}{\kappa^4}H_0^{m}\, ,
\end{equation}
so the assumption for the evolution of the derivative of the
scalar field is the following,
\begin{equation}\label{mainequation}
\dot{\phi}^2=\gamma H(t)^{-m}\, .
\end{equation}
Special cases of the scenario described in Eq.
(\ref{mainequation}) are $\dot{\phi}^2=M^2$ and
$\dot{\phi}^2=\gamma H(t)^2$ (more commonly appearing in the
literature as $\dot{\phi}^2=\delta V(\phi)$) which lead to pure
quasi-de Sitter Hubble rates and to power-law inflation
respectively. These cases will be briefly studied in the end of
the section. The assumption $\dot{\phi}^2 = \gamma H^{-m}$ is
motivated by attractor scaling solutions in k-mouflage gravity, a
class of modified-gravity theories with nonlinear derivative
self-interactions that achieve Vainshtein screening on small
scales, while they allow cosmic acceleration at late
times~\cite{Brax:2014wla}. In these models, when the
high-curvature asymptotic of the kinetic function takes the
power-law form $K(\chi) \sim \chi^{1+1/m}$ (or similar), the
background equations admit exact tracker solutions that satisfy
$\dot{\phi}^2 \propto H^{-m}$. Thus, this kinetic scaling provides
a natural mechanism for slow-roll inflation which is controlled by
the background curvature, unconventionally, in contrast with
standard slow-roll theories of inflation which are controlled by
an ultra-flat potential. Such tracker conditions are also known in
late-time studies, and specifically, the special case $m=2$ is
used in covariant Galileon tracker solutions
\cite{DeFelice:2010aj,Bellini:2012qn}. Therefore, the class of
solutions used in Eq. (\ref{mainequation}) can be viewed as an
early Universe inflationary counterpart of the aforementioned
late-time tracker scaling solutions of scalar-tensor gravity. Let
us proceed with the choice (\ref{mainequation}) and for
consistency, we shall assume that $m>2$ and also it is an even
number $m=2\ell$ or even a rational number of the form
$m=\frac{2\ell}{2 n_1+1}$, with $\ell, n_1$ positive integers.
Thus the assumptions on the parameter $m$ are the following,
\begin{equation}\label{assumptionsonm}
m>2,\,\,\,m=2\ell,\,\,\,\mathrm{or}\,\,\,m=\frac{2\ell}{2
n_1+1},\,\,\ell, n_1=1,2,...\, .
\end{equation}
Now for the choice (\ref{mainequation}), the Raychaudhuri equation
(\ref{fr2}) reads,
\begin{equation}\label{maindiffequation}
2\dot{H}=-\kappa^2\gamma H(t)^{-m}\, ,
\end{equation}
with solution,
\begin{equation}\label{mainsolution}
H(t)=2^{-\frac{1}{m+1}} (m+1)^{\frac{1}{m+1}} \left(2
\omega-\gamma \kappa ^2 t\right)^{\frac{1}{m+1}}\, ,
\end{equation}
with $\omega$ an integration constant with mass dimensions
$[\omega]=[m]^{m+1}$. So we shall assume that $\omega=\lambda\,
\kappa^{-m-1}$ and $\lambda$ is some dimensionless constant. But
we keep the notation with $\omega$ and we will use $\omega=\lambda
\,\kappa^{-m-1}$ later on. Now for consistency $2 \omega-\gamma
\kappa ^2 t>0$ during the inflationary regime and also one must
check explicitly that the evolution of Eq. (\ref{mainsolution})
describes indeed inflation. We shall do that in the end of our
analysis. To proceed, the scale factor corresponding to the Hubble
rate of Eq. (\ref{mainsolution}) is,
\begin{equation}\label{mainscalefactor}
a(t)=a_0 \exp \Big[-\frac{2^{-\frac{1}{m+1}} \,(m+1)^{n}
\,}{\gamma  \kappa ^2 (m+2)}\,\Big(2 \omega-\gamma \kappa ^2
t\Big)^{n}\Big]\, ,
\end{equation}
where we introduced the parameter $n$,
\begin{equation}\label{nparameter}
n=\frac{m+2}{m+1}\, ,
\end{equation}
for reading convenience.  Now the definition of the slow-roll
indices for scalar field inflation studies is
\cite{Hwang:2005hb,Hwang:2001fb,Noh:2000kr,Hwang:2002fp,Noh:2004rt,Hwang:2005hd,Hwang:2006iw},
\begin{equation}\label{definitionslowrollindices}
\epsilon_1=-\frac{\dot{H}}{H^2},\,\,\epsilon_2=\frac{\ddot{\phi}}{H\dot{\phi}}\,
,
\end{equation}
and therefore by combining Eqs. (\ref{maindiffequation}) and
(\ref{mainequation}), it is easy to obtain that, for the case at
hand, we have the general relations,
\begin{equation}\label{generalrelationsslowroll}
\epsilon_1=\frac{\gamma
\kappa^2}{2}H^{-m-2},\,\,\,\epsilon_2=\frac{m}{2}\,\epsilon_1\, .
\end{equation}
Also we can find the analytic solution for the scalar field
$\phi(t)$, by plugging the solution (\ref{mainsolution}) to the
differential equation (\ref{mainequation}), and we obtain,
\begin{equation}\label{mainsolutionscalarfield}
\phi(t)=\phi_0+\frac{2^{\frac{m}{2 m+2}+1}
(m+1)^{n/2}}{\sqrt{\gamma } \kappa ^2 (m+2)}\,\Big(2 \omega-\gamma
\kappa ^2 t\Big)^{n/2}\, ,
\end{equation}
where $\phi_0$ and integration constant, determined by the initial
conditions. We can also find the explicit forms of the slow-roll
indices for the solution (\ref{mainsolution}), which are,
\begin{equation}\label{explicitsolutions1}
\epsilon_1= 2^{\,\frac{1}{m+1}} (m+1)^{-n}\,\gamma \, \kappa ^2
\,\Big(2 \omega-\gamma \kappa ^2 t\Big)^{-n}\, ,
\end{equation}
and
\begin{equation}\label{explicitsolutions2}
\epsilon_2= 2^{-\frac{m}{m+1}} m (m+1)^{-n}\,\gamma \, \kappa ^2\,
\Big(2 \omega-\gamma \kappa ^2 t\Big)^{-n}\, ,
\end{equation}
and we can verify that indeed $\epsilon_2=\frac{\epsilon_1}{2}$.
Now since from Eq. (\ref{generalrelationsslowroll}) we can see
that $\epsilon_1\ll 1$ and also $\epsilon_2\ll 1$, the slow-roll
conditions are justified, so normally the expressions for the
spectral index of the scalar perturbations,
\begin{equation}\label{scalarpreliminary}
n_{\mathcal{S}}=1-4\epsilon_1-2\epsilon_2\, ,
\end{equation}
and for the tensor-to-scalar ratio,
\begin{equation}\label{tensortoscalarpreliminary}
r=16\,\epsilon_1\, ,
\end{equation}
should apply. However, we have to prove that directly from the
power-spectrum. We shall use only the assumption that
$\epsilon_1\ll 1$ and also $\epsilon_2\ll 1$. In addition, the
derivative of the first slow-roll index with respect to the cosmic
time is in the case at hand,
\begin{equation}\label{dotepsilonanalytic}
\dot{\epsilon}_1=\frac{\gamma^2\kappa^2
(m+2)}{4}H^{-2m-3}=\Big(\frac{\gamma
\kappa^2}{2}\Big)^{\frac{2m+5}{m+2}}(m+2)\,\epsilon_1^{n+1}\, ,
\end{equation}
and also $\dot{\epsilon}_2=\frac{m}{2}\dot{\epsilon}_1$. Thus,
since $m>2$, we have $\dot{\epsilon}_1\ll \epsilon_1 \ll 1$ and
$\dot{\epsilon}_2\ll \epsilon_1\ll 1$. We shall use this in the
following subsection. The parameter $\beta$ which appears in the
definition of the parameter $\gamma$ in Eq. (\ref{gammadef}) will
be constrained by the amplitude of the scalar perturbations the
explicit form of which we shall find in the next section. Next, by
using the Friedmann equation (\ref{fr1}), one easily obtains the
scalar field potential as a function of the cosmic time, the
analytic form of which we quote it in the Appendix. For the
potential we found, the scalar field equation of motion
(\ref{eomphi}) is trivially satisfied. We can also express the
scalar potential as a function of the scalar field, by inverting
the solution $\phi(t)$ we obtained in Eq.
(\ref{mainsolutionscalarfield}), so the scalar field potential as
a function of the scalar field, namely $V(\phi)$ is equal to,
\begin{equation}\label{scalarfieldpotentialasfunctionofphi}
V(\phi)=\mathcal{B}\Big{(}3 \kappa ^2 (m+2)^2 (\phi
-\phi_0)^2-\mathcal{K} \Big{)}\Big(\varphi -\phi_0\Big)^{-\frac{2
m}{m+2}}\, ,
\end{equation}
where the constants $\mathcal{K}$ and $\mathcal{B}$ are given in
the Appendix.

\subsection{Cosmological Perturbations and the Observational Indices}

In this section, we shall derive the expressions for the spectral
index of the scalar perturbations and the tensor-to-scalar ratio
for the scalar theory which obeys the differential equation
(\ref{mainequation}). We shall use the notation and presentation
flow of Ref. \cite{Hwang:2005hb} and we refer to that reference
for more details. Also we shall work in a flat FRW background
metric, which we will perturb. The scalar- and tensor-type
perturbations of the FRW metric are the following,
\bea
   & & d s^2 = - a^2 \left( 1 + 2 \alpha \right) d \eta^2
       - 2 a^2 \beta_{,\alpha} d \eta d x^\alpha
       + a^2 \left( g^{(3)}_{\alpha\beta}
       + 2 \varphi g^{(3)}_{\alpha\beta}
       + 2 \gamma_{,\alpha|\beta}
       + 2 C_{\alpha\beta} \right) d x^\alpha d x^\beta,
   \label{metric}
\eea where $a(\eta)$ is the cosmic scale factor expressed in terms
of the conformal time $\eta$ defined as $ dt \equiv a d \eta$. The
variables $\alpha$, $\beta$, $\gamma$ and $\varphi$ are scalar
perturbations depending on the spacetime variables. The tensor
perturbation $C_{\alpha\beta}$ is transverse and trace-free. The
metric $g^{(3)}_{\alpha\beta}$ indicates the comoving background
three-space part of the FRW metric, being spatially homogeneous
and isotropic, with, \bea
   g^{(3)}_{\alpha\beta} d x^\alpha d x^\beta
   &=& {1 \over \left( 1 + \bar r^2 \right)^2}
       \left( d x^2 + d y^2 + d z^2 \right)\,. \eea
Ignoring rotation perturbations, the kinematic quantities in the
normal frame are, \bea
   & & \theta = 3 H , \quad
       \sigma_{\alpha\beta}
       = \chi_{,\alpha|\beta}
       - {1 \over 3} g_{\alpha\beta}^{(3)} \Delta \chi
       + a^2 \dot C^{(t)}_{\alpha\beta}, \quad
       a_\alpha = \alpha_{,\alpha}, \quad
       R^{(h)} = {1 \over a^2} \left[ 6 K - 4 \Delta
       \varphi \right],
   \label{kinematic-quantities}
\eea where \bea
   & & \chi \equiv a \left( \beta + a \dot \gamma \right),
   \label{chi-def}
\eea $\Delta$ is the Laplacian operator for
$g^{(3)}_{\alpha\beta}$ and $\theta$ is the expansion scalar,
$\sigma_{ab}$ is the  shear tensor $a_a$ is the acceleration
vector, and recall that in the normal-frame we have the vanishing
of the rotation vector $\omega_{\alpha\beta} = 0$. For the energy
momentum tensor we have, \bea
   & & T^0_0 = - \left( \bar \mu + \delta \mu \right), \quad
       T^0_\alpha = - \left( \mu + p \right) v_{,\alpha}, \quad
       T^\alpha_\beta = \left( \bar p + \delta p \right) \delta^\alpha_\beta
       + \Pi^\alpha_\beta,
   \label{Tab}
\eea with $\Pi^\alpha_\beta$ being a trace-free anisotropic stress
and $\Pi^\alpha_\beta$ being calculated on
$g^{(3)}_{\alpha\beta}$. The barred quantities are background
metric quantities. We decompose the anisotropic stress in the
following way, \bea
   & & \Pi_{\alpha\beta} \equiv {1 \over a^2} \left( \Pi_{,\alpha|\beta}
       - {1 \over 3} g^{(3)}_{\alpha\beta} \Delta \Pi \right)
       + \Pi_{\alpha\beta}^{(t)},
   \label{Pi}
\eea with $\Pi_{\alpha\beta}^{(t)}$ being transverse and
trace-free. Also the entropic perturbation $e$ is defined as
follows, \bea
   & & e \equiv \delta p - c_s^2 \delta \mu, \quad
       c_s^2 \equiv \dot p / \dot \mu.
\eea Using the gauge transformation, $\hat  x^a \equiv x^a +
\tilde \xi^a (x^e)$ we get, \bea
   & & \hat \alpha = \alpha - \dot \xi^t, \quad
       \hat \beta = \beta - {1 \over a} \xi^t
       + a \left( {\xi \over a} \right)^\cdot, \quad
       \hat \gamma = \gamma - {1 \over a} \xi, \quad
       \hat \varphi = \varphi - H \xi^t, \quad
       \hat \chi = \chi - \xi^t, \quad
       \hat \kappa = \kappa
       + \left( 3 \dot H + {\Delta \over a^2} \right) \xi^t,
   \nonumber \\
   & & \delta \hat \mu = \delta \mu - \dot \mu \xi^t, \quad
       \delta \hat p = \delta p - \dot p \xi^t, \quad
       \hat v = v - {1 \over a} \xi^t, \quad
       \hat \Pi = \Pi, \quad
       \delta \hat \phi = \delta \phi - \dot \phi \xi^t; \quad
       \hat C_{\alpha\beta} = C_{\alpha\beta}, \quad
       \hat \Pi_{\alpha\beta}^{({t})} = \Pi_{\alpha\beta}^{({t})},
   \label{GT}
\eea with $\xi^0 \equiv {1 \over a} \xi^t$ and $\xi_\alpha \equiv
\xi_{,\alpha}$. Also $\bar \phi$ and $\delta \phi$ are the
background and the perturbation part of the scalar field
$\phi({\bf x}, t)$. We shall use the following several
gauge-invariant  variables, \bea
   & & \varphi_\chi \equiv \varphi - H \chi, \quad
       \varphi_v \equiv \varphi - a H v, \quad
       \delta_v \equiv \delta - a {\dot \mu \over \mu} v, \quad
       \delta \phi_\varphi \equiv \delta \phi
       - {\dot \phi \over H} \varphi
       \equiv - {\dot \phi \over H} \varphi_{\delta \phi}, \quad
       v_\chi \equiv v - {1 \over a} \chi
       \equiv - {1 \over a} \chi_v,
   \label{GI-variables}
\eea with $\delta \equiv \delta \mu / \mu$. The gauge-invariant
variable $\delta \phi_\varphi$ is equivalent to $\delta \phi$ in
the curvature gauge, which has the gauge condition $\varphi \equiv
0$. For the scalar field case we have, \bea
   & & \mu \equiv {1 \over 2} \dot \phi^2 + V, \quad
       p \equiv {1 \over 2} \dot \phi^2 - V,
   \label{MSF-fluid-BG} \\
   & & \delta \mu \equiv \dot \phi \delta \dot \phi
       - \dot \phi^2 \alpha + V_{,\phi} \delta \phi, \quad
       \delta p \equiv \dot \phi \delta \dot \phi
       - \dot \phi^2 \alpha - V_{,\phi} \delta \phi, \quad
       v \equiv {1 \over a} { \delta \phi \over \dot \phi },
       \quad
       \Pi = 0 = \Pi^{(t)\alpha}_{\;\;\;\; \beta},
   \label{MSF-fluid-pert}
\eea with $\phi ({\bf x}, t) = \bar \phi (t) + \delta \phi ({\bf
x}, t)$. In addition, the background and perturbed scalar field
equations of motion are, \bea
   & & \ddot \phi + 3 H \dot \phi + V_{,\phi} = 0,
   \\
   & & \delta \ddot \phi + 3 H \delta \dot \phi
       - {\Delta \over a^2} \delta \phi + V_{,\phi\phi} \delta \phi
       = \dot \phi \left(  \dot \alpha \right)
       + \left( 2 \ddot \phi + 3 H \dot \phi \right) \alpha.
   \label{EOM-MSF-pert}
\eea Since $\delta \phi = 0$ yields $v = 0$, the uniform-field
gauge ($\delta \phi \equiv 0$) is identical to the comoving gauge
($v \equiv 0$), hence $\varphi_v = \varphi_{\delta \phi}$.
Therefore, we get, \bea
   & & \Phi \equiv \varphi_{\delta \phi},
   \label{Phi-def-MSF}
\eea thus, \bea
   & & \dot \Phi = {H \over 4 \pi G \dot \phi^2 } {c_A^2 \Delta \over a^2}
       \varphi_\chi,
   \label{dot-Phi-eq-MSF} \\
   & & {H \over a} \left( {a \over H} \varphi_\chi \right)^\cdot
       = {4 \pi G \dot \phi^2 \over H} \Phi,
   \label{dot-Psi-eq-MSF}
\eea with \bea
   & & c_A^2 \equiv 1, \quad
       c_s^2 \equiv {\dot p \over \dot \mu}
             = -1 - {2 \ddot \phi \over 3 H \dot \phi}.
\eea Thus in the end, the perturbation equations are written as
follows, \bea
   & & {H^2 c_A^2 \over a^3 \dot \phi^2}
       \left[ {a^3 \dot \phi^2 \over H^2 c_A^2} \dot \Phi \right]^\cdot
       = c_A^2 {\Delta \over a^2} \Phi,
   \label{ddot-Phi-eq-MSF} \\
   & & {\dot \phi^2 \over H} \left[ {H^2 \over a \dot \phi^2}
       \left( {a \over H} \varphi_\chi \right)^\cdot \right]^\cdot
       = c_A^2 {\Delta \over a^2} \varphi_\chi,
   \label{ddot-Psi-eq-MSF}
\eea where $\Delta = - k^2$, $k$ is the wavenumber in Fourier
space, and in the case of a canonical minimally coupled scalar
field in flat FRW, $c_A=1$ in natural units, and also $c_A$ is the
sound wave speed of the scalar field and of the perturbed metric.
We can further write the scalar perturbations in more convenient
format in the form of Mukhanov-Sasaki equation, by introducing the
following variables, \bea
   & & z= \frac{a\dot{\phi}}{H}, \quad
    \quad
       \tilde v \equiv z \Phi, \quad
       u \equiv \frac{1}{\kappa^2}\frac{a}{H} {1 \over z} \Psi,
   \label{u-v-def}
\eea and the perturbation equations are, \bea
   & & \tilde v^{\prime\prime}
       - \left(\Delta + {z^{\prime\prime} \over z} \right) \tilde v = 0,
   \label{v-eq} \\
   & & u^{\prime\prime} - \left[ \Delta
       + {(1/\bar z)^{\prime\prime} \over (1/\bar z)} \right] u= 0.
   \label{u-eq}
\eea where recall that we took $ c_A^2=1$ for the minimally
coupled scalar field in the FRW spacetime. For the tensor modes we
introduce, \bea
   & & z_t \equiv \frac{a}{\kappa}, \quad
       v_t \equiv z_t \Phi,
   \label{z-v-def-GW}
\eea where $\Phi = C_{\alpha\beta}$, and thus we have the
Mukhanov-Sasaki equation for the tensor perturbations, \bea
   & & v_t^{\prime\prime}
       - \left(  \Delta + {z_t^{\prime\prime} \over z_t} \right) v_t
       =0\, ,
   \label{v-eq-GW}
\eea and we took the speed of gravitational tensor modes
perturbations to be $c_T^2=1$ in natural units for a minimally
coupled scalar field. The perturbation equations have exact
solutions, if the wave speed is constant and if $z \propto
|\eta|^{q}$. In this case, we have \bea \label{auxiliary1}
   & & {z^{\prime\prime} \over z}
       = {q(q-1) \over \eta^2} \equiv {n \over \eta^2},
\eea and the exact solutions for the perturbations are \bea
   & & \Phi (k, \eta) = {\sqrt{ \pi |\eta|} \over 2 z}
       \left[ c_1 ({k}) H_\nu^{(1)} (c_A k |\eta|)
       + c_2 ({k}) H_\nu^{(2)} (c_A k |\eta|) \right],
   \label{Phi-exact-sol} \\
   & & \Psi (k, \eta) = - {\sqrt{ \pi |\eta|} \over 2 z}
       {a c_A \over 2 k x_1}
       \left[ c_1 ({k}) H_{\nu - 1}^{(1)} (c_A k |\eta|)
       + c_2 ({k}) H_{\nu - 1}^{(2)} (c_A k |\eta|) \right],
   \label{Psi-exact-sol}
\eea with \bea
   & & \nu \equiv {1 \over 2} - q = \sqrt{ n + {1 \over 4}}.
\eea Now to quantize the field perturbations, we use the perturbed
action \bea
   \delta^2 S
   &=& {1 \over 2} \int a z^2 \left( \dot \Phi^2
       - {1 \over a^2} \Phi^{,\alpha} \Phi_{,\alpha} \right) dt d^3 x,
   \nonumber \\
   &=& {1 \over 2} \int \left( \tilde v^{\prime 2}
       - \tilde v^{,\alpha} \tilde v_{,\alpha}
       + {z^{\prime\prime} \over z} \tilde v^2 \right)
       d \eta d^3 x.
   \label{perturbed-action}
\eea We shall expand the field in Fourier modes, using a canonical
quantum field theory approach, \bea
   & & \hat \Phi ( {\bf x}, t)
       \equiv \int {d^3 k \over ( 2 \pi)^{3/2} }
       \left[ \hat a_{\bf k} \Phi_{\bf k} (t) e^{i {\bf k}\cdot {\bf x}}
       + \hat a^\dagger _{\bf k} \Phi^*_{\bf k} (t) e^{-i
       {\bf k} \cdot {\bf x}} \right],
   \nonumber \\
   & & [ \hat a_{\bf k} , \hat a_{{\bf k}^\prime} ] = 0, \quad
       [ \hat a^\dagger_{\bf k} , \hat a^\dagger_{{\bf k}^\prime} ] = 0, \quad
       [ \hat a_{\bf k} , \hat a^\dagger_{{\bf k}^\prime} ]
       = \delta^3 ( {\bf k} - {\bf k}^\prime ).
\eea with the conjugate momentum being $\pi_\Phi \equiv {\partial
{\cal L} \over \partial \dot \Phi}
     = a z^2 \dot \Phi$.
The isochronous quantization relations are, $[ \hat \Phi ({\bf
x},t), {\hat \pi}_\Phi ({\bf x}^\prime, t) ]
     = i \delta^3 ({\bf x} - {\bf x}^\prime)$
becomes $[ \hat \Phi ({\bf x},t), \dot {\hat \Phi} ({\bf
x}^\prime, t) ]
     = {i \over a z^2} \delta^3 ({\bf x} - {\bf x}^\prime)$
and in effect we have the Wronskian condition \bea
   & & \Phi_{\bf k} \dot \Phi_{\bf k}^*
       - \Phi_{\bf k}^* \dot \Phi_{\bf k}
       = {i \over a z^2}.
   \label{Phi-quantization}
\eea For $z \propto |\eta|^{q}$, and using Eqs.
(\ref{Phi-exact-sol})  and (\ref{Psi-exact-sol}), the Fourier
modes $\Phi_k(\eta)$ and $\Psi_k(\eta)$, take the form, \bea
   & & \Phi_{\bf k} (\eta) = {\sqrt{ \pi |\eta|} \over 2 z}
       \left[ c_1 ({k}) H_\nu^{(1)} (c_A k |\eta|)
       + c_2 ({k}) H_\nu^{(2)} (c_A k |\eta|) \right],
   \label{Phi-mode-sol} \\
   & & \Psi_{\bf k} (\eta) = - {\sqrt{ \pi |\eta|} \over 2 z}
       {a c_A \over 2 k x_1}
       \left[ c_1 ({k}) H_{\nu - 1}^{(1)} (c_A k |\eta|)
       + c_2 ({k}) H_{\nu - 1}^{(2)} (c_A k |\eta|) \right],
   \label{Psi-mode-sol}
\eea with \bea
   & & |c_2 ({k})|^2 - |c_1 ({k})|^2 = 1,
\eea which follows from the quantization condition in eq.
(\ref{Phi-quantization}). The power-spectrum of the perturbations
is, \bea
   & & {\cal P}_{\hat \Phi} (k,t)
       \equiv {k^3 \over 2 \pi^2} \int
       \langle \hat \Phi ({\bf x} + {\bf r}, t)
       \hat \Phi ({\bf x}, t) \rangle_{\rm vac} e^{-i {\bf k} \cdot {\bf r}}
       d^3 r
       = {k^3 \over 2 \pi^2} | \Phi_k (t) |^2.
   \label{P-vacuum}
\eea so using (\ref{Phi-mode-sol}) we get the resulting expression
for the power spectrum,
 \bea
   & & {\cal P}^{1/2}_{\hat \Phi} ({\bf k}, \eta)
       = {H \over 2 \pi}
       {1 \over a H |\eta|}
       {\Gamma (\nu) \over \Gamma (3/2)}
       \left( {k |\eta|\over 2} \right)^{3/2 -\nu}
       {1 \over  z/a}.
   \label{P-Phi}
\eea Also the tensor power spectrum is, \bea
   & & {\cal P}^{1/2}_{\hat C_{\alpha\beta}} ({\bf k}, \eta)
       = \sqrt{16 \pi G} {H \over 2 \pi} {1 \over a H |\eta|}
       {\Gamma (\nu_t) \over \Gamma (3/2)}
       \left( {k |\eta|\over 2} \right)^{3/2 -\nu_t}
       {1 / \sqrt{8 \pi G} \over  z_t/a}.
   \label{P-GW}
\eea Now the spectral index of the scalar perturbations is defined
as follows,
\begin{equation}\label{scalarspectralindex}
 n_{\mathcal{S}} - 1= {\partial \ln{{\cal P}_\Phi} \over \partial \ln{k}}\, ,
\end{equation}
and the tensor spectral index is accordingly defined, thus we
have,
 \bea
   n_{\mathcal{S}} - 1= 3 - 2 \nu = 2 + 2 q.
   \label{n_ST}
\eea We now show that if the slow-roll indices $\epsilon_1$ and
$\epsilon_2$ in the case of single scalar field theory are
$\epsilon_i\ll 1$, then we can cast $z''/z$ in the form of Eq.
(\ref{auxiliary1}). Indeed, we have,
 \bea
   {z^{\prime\prime} \over z}
   &=& a^2 \Bigg[
       H^2 \left( 1 + \epsilon_1 + \epsilon_2 \right)
       \left( 2 + \epsilon_2 \right)
       + H \left(  \dot \epsilon_1 + \dot \epsilon_2 \right) \Bigg],
   \label{z-over-z-S} \\
   {z_t^{\prime\prime} \over z_t}
   &=& a^2 H^2 \left( 2 - \epsilon_1  \right).
   \label{z-over-z-T}
\eea For the moment we did not use $\epsilon_i\ll 1$, but now we
assume that $\epsilon_i\ll 1$ and also in Eq.
(\ref{dotepsilonanalytic}) we proved that $\dot{\epsilon}_1\ll
\epsilon_1 \ll 1$ and $\dot{\epsilon}_2\ll \epsilon_1\ll 1$. Thus,
at leading order we have \cite{Oikonomou:2020krq},
\begin{equation}\label{etahorizoncrossing}
\eta=-\frac{1}{aH}\frac{1}{1-\epsilon_1}\, .
\end{equation}
In Ref. \cite{Hwang:2005hb} made a crucial assumption, namely that
$\dot{\epsilon}_1=0$, and there is also an alternative version in
the literature,  $\epsilon_1=$const \cite{Stewart:1993bc}. As we
proved though in \cite{Oikonomou:2020krq}, only the requirement
$\dot{\epsilon}_i\ll 1$ is needed, in order for Eq.
(\ref{etahorizoncrossing}) to be valid. Hence, in view of Eq.
(\ref{etahorizoncrossing}), Eqs.
(\ref{z-over-z-S},\ref{z-over-z-T}) take the form \bea
    {z^{\prime\prime} \over z}
   &=& {1 \over \eta^2} {1\over ( 1 - \epsilon_1 )^2 }
       \left( 1 + \epsilon_1 + \epsilon_2 \right)
       \left( 2 + \epsilon_2 \right)
       \equiv {n_s \over \eta^2},
   \\
   {z_t^{\prime\prime} \over z_t}
   &=& {1 \over \eta^2} { (2 - \epsilon_1 )
       \over ( 1 + \epsilon_1 )^2 }
       \equiv {n_t \over \eta^2}\,
\eea which have exactly the form as in Eq. (\ref{auxiliary1}).
Thus, the scalar and tensor spectral indices become, \bea
\label{spectralindices}
   & & n_{\mathcal{S}} - 1 = -4\epsilon_1-2\epsilon_2,
   \\
   & & n_T = 3 - 2 \nu_t = 3 - \sqrt{4 n_t + 1}
       = -2\epsilon_1 \, ,
\eea and accordingly, the tensor-to-scalar ratio reads,
\begin{equation}\label{tensortoscalarrationfinal}
r=16\epsilon_1\, .
\end{equation}
The amplitude of the scalar perturbations can easily be obtained
and it has the following form,
\begin{equation}\label{amplitude}
\mathcal{P}_{\hat \Phi}(k)=\frac{H^4}{4\pi^2
\dot{\phi}^2}\Big{|}_{k=a\,H}\, ,
\end{equation}
evaluated at first horizon crossing of the mode $k$, when
$k=a\,H$. Thus we proved that in the case in which the scalar
field satisfies the condition (\ref{mainequation}), with $m>2$,
the spectral index of the scalar perturbations and the
tensor-to-scalar ratio are given as in Eq.
(\ref{scalarpreliminary}) and (\ref{tensortoscalarpreliminary}),
which are the usual expressions in scalar field theory. Now in the
next subsection we shall analyze the phenomenology of the scalar
field theories which satisfy the condition (\ref{mainequation}).

\subsection{Phenomenology of the Analytic Model of Inflation}

\begin{figure}
\centering
\includegraphics[width=28pc]{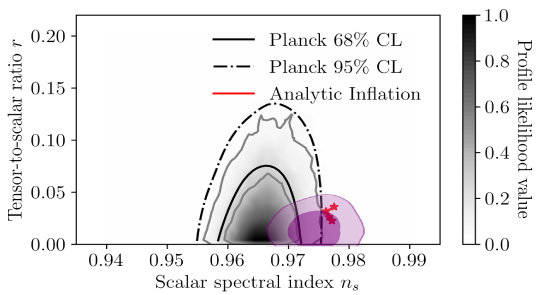}
\caption{Marginalized curves of the Planck 2018 data and the
analytic slow-roll model (\ref{mainequation}), confronted with the
ACT data, the Planck 2018 data, and the updated Planck constraints
on the tensor-to-scalar ratio for $N\sim 50-60$ and $m$ in the
range $m=[6,14]$.}\label{plot1}
\end{figure}

Now let us analyze the inflationary phenomenology of the model
(\ref{mainequation}), which leads to the Hubble rate
(\ref{mainsolution}) and the scalar field
(\ref{mainsolutionscalarfield}). The phenomenology can be easily
analyzed because we have obtained analytic relations for all the
quantities involved. Let us start by calculating the time instance
$t_f$ at which the inflationary era ends. This is obtained by
solving the equation $\epsilon_1(t_f)=1$, and so we obtain,
\begin{equation}\label{tf}
t_f=\frac{2 \omega}{\gamma \kappa ^2}-\frac{2^{\frac{1}{m+2}}\,
\gamma ^{1/n} \kappa ^{2/n}}{(m+1)\,\gamma  \kappa ^2}\, ,
\end{equation}
where $n$ is defined in Eq. (\ref{nparameter}). Also we can find
the time of the first horizon crossing, $t_i$, by using the
definition of the $e$-foldings number,
\begin{equation}\label{efoldingsnumberdef}
N=\int_{t_i}^{t_f}H(t)\mathrm{d}t\, ,
\end{equation}
so we get,
\begin{equation}\label{ti}
t_i=\frac{2 \omega}{\gamma  \kappa ^2}-\frac{\Big[\gamma  \kappa
^2 2^{\frac{1}{m+1}} (m+1)^{-n} (m+2) \Big(N-\Gamma
\Big)\Big]^{1/n}}{\gamma \kappa ^2}\, ,
\end{equation}
where for convenience we introduced the parameter $\Gamma$ which
is given in the Appendix too. The first slow-roll index in terms
of the cosmic time is given in Eq. (\ref{explicitsolutions1}) and
recall we found that $\epsilon_2=\frac{m}{2}\epsilon_1$. Now we
can evaluate the first slow-roll index as a function of the
$e$-foldings number and it takes a very simple form,
\begin{equation}\label{firstslowrollfinal}
\epsilon_1=\frac{1}{1+(m+2) N}\, .
\end{equation}
Thus the scalar spectral index (\ref{scalarpreliminary}) also
takes a very simple form and it reads,
\begin{equation}\label{scalarasfunctionofm}
n_{\mathcal{S}}=1-\frac{m+4}{1+(m+2) N}\, ,
\end{equation}
while the tensor-to-scalar ratio (\ref{tensortoscalarpreliminary})
reads,
\begin{equation}\label{tensortoscalarratio}
r=\frac{16}{1+(m+2) N}\, .
\end{equation}
The resulting inflationary phenomenology depends only on one
parameter and this is a particularly interesting result, since the
observational indices depend solely on the parameter $m$ and not
on the parameter $\beta$ appearing implicitly in Eq.
(\ref{mainequation}). Let us find the leading order expressions of
the spectral index and of the tensor-to-scalar ratio, so in the
large $N$ limit, these read,
\begin{equation}\label{nslargeN}
n_{\mathcal{S}}\simeq 1-\frac{m+4}{(m+2) N}+\frac{m+4}{(m+2)^2
N^2}\, ,
\end{equation}
\begin{equation}\label{rlargeN}
r\simeq \frac{16}{(m+2) N}-\frac{16}{(m+2)^2 N^2}\, .
\end{equation}
We can easily investigate the phenomenology of the model for the
various values of the parameter $m$ which recall that it is
constrained as in Eq. (\ref{assumptionsonm}). Let us give some
examples by choosing the $e$-foldings number in the range $N\sim
50-60$. So for example if $m=6$ and $N=55.5$ we get,
$n_{\mathcal{S}}=0.977528$ and $r=0.0359549$ which are compatible
with the ACT and the updated BICEP/Planck data. Generally, for
compatibility with the ACT and the updated BICEP/Planck data we
need $m\geq 6$. Note that the current model cannot yield a
spectral index compatible with the Planck data, for any allowed
value of the parameter $m$. So this model is suited perfectly only
for the ACT-compatible inflationary theories. Other characteristic
set of $(m,N)$ values compatible with the ACT and updated
BICEP/Planck data are, $m=8$ and $N=50.5$ we get,
$n_{\mathcal{S}}=0.976285$ and $r=0.0316204$, or for example
$m=10$ and $N=50$ we get, $n_{\mathcal{S}}=0.976706$ and
$r=0.0263591$, or for $m=14$ and $N=50$ we get,
$n_{\mathcal{S}}=0.977176$ and $r=0.022598$. In Fig. \ref{plot1}
we confront the phenomenology of the model for various $N$ values
in the range $N\sim 50-60$ and $m$ in the range $m=[6,14]$ and as
it can be seen the model is compatible with the ACT and the
updated Planck data. Now the question is, does the parameter $m$
affect the slow-roll conditions? For example if $m$ gets extremely
large values, does it make the model incompatible with the
slow-roll conditions? We can easily examine this by expanding the
first slow-roll index (\ref{firstslowrollfinal}) for very large
$m$ values. In fact a particularly interesting behavior occurs in
the observational indices in the large $m$ limit, and
specifically, the spectral index in the large $N$ limit reads,
\begin{equation}\label{largeNns}
n_{\mathcal{S}}\simeq 1-\frac{1}{N}+\frac{1-2 N}{m
N^2}+\mathcal{O}\Big(\frac{1}{m^3}\Big)\, ,
\end{equation}
so it is basically independent of the parameter $m$ and behaves as
$n_{\mathcal{S}}\simeq 1-\frac{1}{N}$. On the other hand the
tensor-to-scalar ratio behaves as,
\begin{equation}\label{asymptoticm}
r\sim \frac{16}{m N}-\frac{16 (2 N+1)}{m^2
N^2}+\mathcal{O}\Big(\frac{1}{m^3}\Big)\, .
\end{equation}
Thus for $N\sim 50-60$ the spectral index takes values in the
range $N\sim [0.98-0.98333]$ and the tensor-to-scalar ratio takes
particularly small values, since it strongly depends on the
parameter $m$. On the other hand, the spectral index does not
depend on the parameter $m$ and thus its behavior is regulated by
the $e$-foldings number. Notice that the model at hand at large
$m$ yields a very small tensor-to-scalar ratio and a spectral
index of the order $n_{\mathcal{S}}\sim 0.98$ way more blue in
comparison to the Planck acceptable values. This result can be
useful for future searches that may validate the ACT data, or even
find values in tension with even the ACT or Planck data.
Another interesting behavior of the one-parameter inflationary
model we developed is the functional dependence of the spectral
index as a function of the tensor-to-scalar ratio. This is easily
obtained, since $\epsilon_1=\frac{m}{2}\epsilon_2$ and
$r=16\epsilon_1$, so we get,
\begin{equation}\label{nsrrelation}
n_{\mathcal{S}}=1-\frac{r}{4}-\frac{m}{16}\, r\, .
\end{equation}
This is an exact relation and it shows one thing, that the model
produces a universal curve in the $n_s-r$ plane, which makes it
extremely easy to comply with ACT/Planck/BICEP constraints.

Now let us consider several other phenomenological features of the
analytic slow-roll model of inflation. Let us first consider the
running of the spectral index, which is defined as,
\begin{equation}\label{runningdef}
a_s=\frac{\mathrm{d} n_{\mathcal{S}}}{\mathrm{d} \ln k}\, ,
\end{equation}
with $k$ being the comoving wavenumber corresponding to a
primordial inflationary mode. We can write the running of the
spectral index $a_s$ in the following form,
\begin{equation}\label{runningdef1}
a_s=\frac{\mathrm{d} n_{\mathcal{S}}}{\mathrm{d} \ln
k}=\frac{\mathrm{d}
n_{\mathcal{S}}}{\mathrm{d}N}\frac{\mathrm{d}N}{\mathrm{d} \ln
k}\, ,
\end{equation}
where $N$ is the $e$-foldings number. In addition, by using the
formula $\frac{\mathrm{d}N}{\mathrm{d} \ln
k}=\frac{1}{1-\epsilon_1}$, we can obtain the following simple
formula for the running of the spectral index $a_s$,
\begin{equation}\label{runningdefmainfinal}
a_s=\frac{1}{1-\epsilon_1}\frac{\mathrm{d}
n_{\mathcal{S}}}{\mathrm{d}N}\, .
\end{equation}
So for the model at hand, by using Eqs.
(\ref{runningdefmainfinal}), (\ref{firstslowrollfinal}) and
(\ref{scalarasfunctionofm}), we have,
\begin{equation}\label{runningfinalmodelanalytic}
a_s=\frac{m+4}{N \Big(1+(m+2) N\Big)}\, .
\end{equation}
A striking first feature is that the running of the spectral index
is positive, in contrast to well-known inflationary models, like
for example the Starobinsky or the Higgs model. It is interesting
to note that the ACT data point towards a positive running of the
spectral index,
\begin{equation}\label{actrunning}
\frac{\mathrm{d}n_{\mathcal{S}}}{\mathrm{d}\ln k}=0.0062 \pm
0.0052\,\,\,(68\%)\, ,
\end{equation}
but with a very small statistic significance $1.2\sigma$. Thus if
this statistical preference of inflationary models with positive
running of the spectral index becomes stronger in the future, the
model at hand might be one of the few models in the context of
general relativity that can produce a positive running. Let us
compute some characteristic values, so for $(m,N)=(6,55.5)$ we
get, $a_s=0.000404899$, while for $(m,N)=(8,50.5)$ we get,
$a_s=0.000469612$. Also for $(m,N)=(10,50)$ we get,
$a_s=0.00046589$ and for $(m,N)=(14,50)$ we get,
$a_s=0.000449438$. All these values are well fitted within the ACT
data, and we used the same $(m,N)$ pairs that we used to examine
the compatibility of the spectral index and the tensor-to-scalar
ratio with the ACT/BICEP/Planck data. Let us also find the
asymptotic behavior of the running of the spectral index at large
$m$ values, and it reads,
\begin{equation}\label{asymptoticas}
a_s\simeq \frac{1}{N^2}+\frac{1-4 N^2}{m^2 N^4}+\frac{2 N-1}{m
N^3}+\mathcal{O}\Big(\frac{1}{m^3}\Big)\, ,
\end{equation}
thus the large $m$ values do not affect the viability of the
running of the spectral index, since for $N\sim 50-60$, the
running of the spectral index is in the range $a_s^{asympt}\sim
[0.0004,0.000277778]$, which are well fitted within the ACT
constraints.

It is useful to find the values of the effective equation of state
(EoS) parameter at the end of inflation (at $N=0$) and at the
beginning of the inflationary era (at $N\sim 50-60$). We have,
\begin{equation}\label{EoS}
w_{eff}=-1-\frac{2}{3}\frac{\dot{H}}{H^2}=-1+\frac{r}{24}\, ,
\end{equation}
with $r$ given in Eq. (\ref{rlargeN}), and $w_{eff}$ must have the
value $w_{eff}=-1/3$ at the end of inflation (at $N=0$) while it
must be near $w_{eff}=-1$ at the beginning of inflation (at $N\sim
50-60$). Indeed, for $N=0$, the EoS parameter is
$w_{eff}=-\frac{1}{3}$ and it is independent of the value of $m$
as it can easily be checked. Also the value of the EoS parameter
at first horizon crossing, for $(m,N)=(6,55.5)$ is
$w_{eff}=-0.998502$, while for $(m,N)=(8,50.5)$ we get,
$w_{eff}=-0.998682$. Also for $(m,N)=(10,50)$ we get,
$w_{eff}=-0.998891$ and for $(m,N)=(14,50)$ we get,
$w_{eff}=-0.999168$. Thus, as $m$ increases, the EoS parameter
approaches the quasi-de Sitter value $w_{eff}=-1$. This is
particularly interesting and can easily be validated by taking the
asymptotic form of Eq. (\ref{EoS}) in the large $m$ regime, and we
have,
\begin{equation}\label{largemweff}
w_{eff}^{asympt}\simeq -1+\frac{2}{3 m N}-\frac{2 (2 N+1)}{3 m^2
N^2}\, .
\end{equation}
This is reasonable, since the main assumption of this work
materialized by Eq. (\ref{mainequation}), in the limit $m\to
\infty$ yields a pure slow-roll evolution, with $w_{eff}\sim -1$
and a de-sitter regime with $\dot{\phi}\sim 0$. The behavior
$w_{eff}\sim -1$ is allegedly attributed to a quasi-de Sitter
evolution, however, we shall prove in a later section that the
quasi-de Sitter evolution is obtained by the condition
$\dot{\phi}^2=\kappa^{-4}\delta $, where $\delta$ is an arbitrary
dimensionless parameter.

Now let us focus on the amplitude of the scalar perturbations,
which we proved that in the case at hand, it is given by Eq.
(\ref{amplitude}) evaluated at the first horizon crossing. So in
view of Eq. (\ref{mainequation}), the amplitude of the scalar
perturbations at first horizon crossing (when the CMB pivot scale
exits the Hubble horizon during inflation at $N_*\sim 50-60$)
takes the final form,
\begin{equation}\label{finalamplitude1}
A_s=\frac{H_*^{m+4}}{4 \pi ^2 \gamma }\, ,
\end{equation}
so when evaluated at the pivot scale, we get,
\begin{equation}\label{amplitudefinal}
A_s=\frac{2^{-\frac{3 m+8}{m+2}} \beta ^{\frac{2}{m+2}} \Big((m+2)
N_*+1\Big)^{\frac{m+4}{m+2}} \tau ^{\frac{2 m}{m+2}}}{\pi ^2}\, ,
\end{equation}
where we introduced the notation,
\begin{equation}\label{Hotau}
H_0=\frac{\tau}{\kappa}\, ,
\end{equation}
with $\tau$ being a dimensionless parameter, and also we used the
definition of $\gamma$ in Eq. (\ref{gammadef}). Also we shall use
the fact that $\omega=\lambda \kappa^{-m-1}$. Now we can evaluate
the constraints on the parameter $\beta$ and $\tau$ that may yield
an inflationary theory compatible with the Planck data. Notice
that the observational indices do not depend on $\tau$ and
$\beta$. In addition, we may examine the behavior of the scale
factor and of the Hubble rate as a function of the cosmic time.
The resulting picture is particularly interesting. For simplicity
let us choose only one $m$ for the presentation. The
generalization is easy. In order to make the plots easier, due to
the huge numbers appearing in the exponential of the scale factor,
we shall rescale the cosmic time as follows:
\begin{equation}\label{rescaledcosmictime}
t=\eta \kappa\, ,
\end{equation}
with $\eta$ being just an arbitrary dimensionless parameter, not
to be confused with the conformal time. In terms of the $\eta$
parameter, the final cosmic time $\eta_f$ at which inflation ends
is $\eta_f=\frac{t_f}{\kappa}$, and specifically,
\begin{equation}\label{etaf}
\eta_f=\frac{2 \lambda  \tau ^{-m}}{\beta }-\frac{
2^{\frac{1}{m+2}}\, (m+2)\, \beta ^{-\frac{1}{m+2}}}{(m+1)^2}\, ,
\end{equation}
and the time instance $\eta_i$ that inflation begins is,
\begin{equation}\label{etainitial}
\eta_i=\frac{ \tau ^{-m} \left(2 \lambda (m+1)-2^{\frac{1}{m+2}}
(m+2)^{\frac{m+1}{m+2}} \beta ^{\frac{m+1}{m+2}} \tau ^{\frac{m
(m+1)}{m+2}} \left(\frac{\tau ^{-m} \left(\left(m^2+3 m+2\right) N
\tau ^m+(m+1)^{\frac{1}{m+1}} (m+2)^{\frac{m+2}{m+1}} \tau
^{\frac{m (m+2)}{m+1}}\right)}{(m+1)
(m+2)}\right)^{\frac{m+1}{m+2}}\right)}{\beta  (m+1)}\, .
\end{equation}
Setting $a_0=1$ in the scale factor (\ref{mainscalefactor}), the
scale factor in terms of $\eta$ reads,
\begin{equation}\label{scalefactoreta}
a(\eta)=\exp \left(-\frac{2^{-\frac{1}{m+1}}
(m+1)^{\frac{1}{m+1}+1} \tau ^{-m} \left(2 \lambda -\beta  \eta
\tau ^m\right)^{\frac{m+2}{m+1}}}{\beta  (m+2)}\right)\, ,
\end{equation}
where we used the fact that $\omega=\lambda \kappa^{-m-1}$. The
physical picture is clear, the Universe starts from an
non-singular point, at $t=0$, and remarkably in the context of
this model, there is no initial singularity. As the cosmic time
grows, the slow-roll inflationary regime starts at $\eta=\eta_i$
($t_i$) and ends at $\eta=\eta_f$ ($t_f$). The Universe reaches
the critical time $\eta_T=\frac{2 \lambda  \tau ^{-m}}{\beta }$,
at which point the scale factor takes the maximum value, but the
Hubble rate becomes equal to zero, since recall that the Hubble
rate in terms of $\eta$ is expressed as follows,
\begin{equation}\label{hubbleasfunctioeta}
H(\eta)=2^{-\frac{1}{m+1}} (m+1)^{\frac{1}{m+1}} \Big(2 \lambda
-\beta \eta  \tau ^m \Big)^{\frac{1}{m+1}}\, .
\end{equation}
Now this is reminiscent of a bounce, but it is not. It is a
turnaround cosmological evolution, with a pressure singularity at
the critical point $\eta_T=\frac{2 \lambda  \tau ^{-m}}{\beta }$,
because the derivative of the Hubble rate diverges severely in
that point. Thus, the energy density and the scale factor are
finite for this cosmological evolution, but the derivative of the
Hubble rate diverges. This is a typical pressure finite-time
singularity according to the Nojiri-Odintsov-Tsujikawa
classification \cite{Nojiri:2005sx}. If the evolution remains
classical, after the turnaround point, the scale factor decreases
and the Universe collapses. However, we will show in a later
section, that when the singularity is reached, quantum phenomena
occur and the conformal anomaly due to conformal particles
dominates the evolution, rendering the classical solution
inactive, and at the same time, the Universe becomes radiation
dominated due to severe particle production, thus offering a
theory for reheating without scalar oscillations and numerous
couplings of the inflaton to the Standard Model particles. Also
thermal effects by the horizon of the finite-time singularity may
also shift the classical evolution from being singular, since once
the singularity is approached, these phenomena initiate and render
the pre-turnaround point classical evolution inactive. These are
very interesting physical phenomena and will be analyzed in depth
in a later section. At this point, we shall be interested in the
classical evolution without taking into account the quantum
phenomena or the thermal phenomena near the singularity.
\begin{figure}
\centering
\includegraphics[width=20pc]{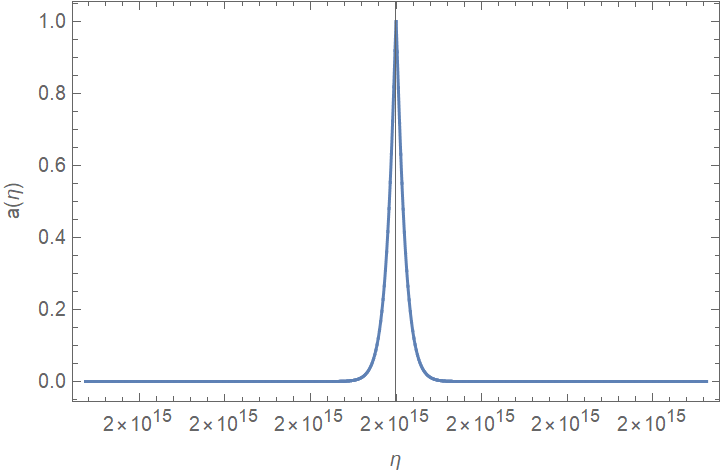}
\includegraphics[width=20pc]{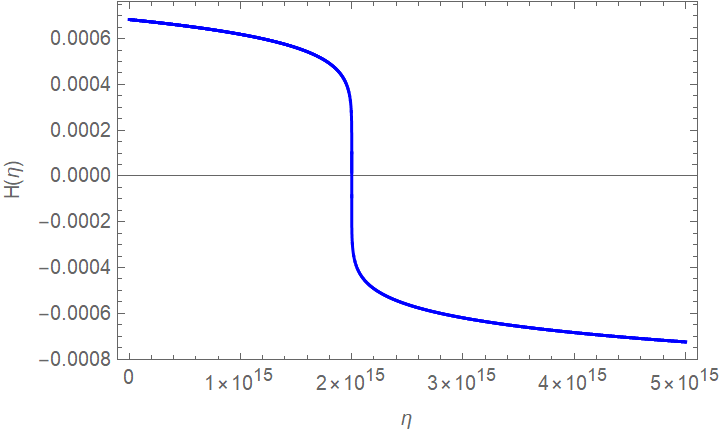}
\includegraphics[width=20pc]{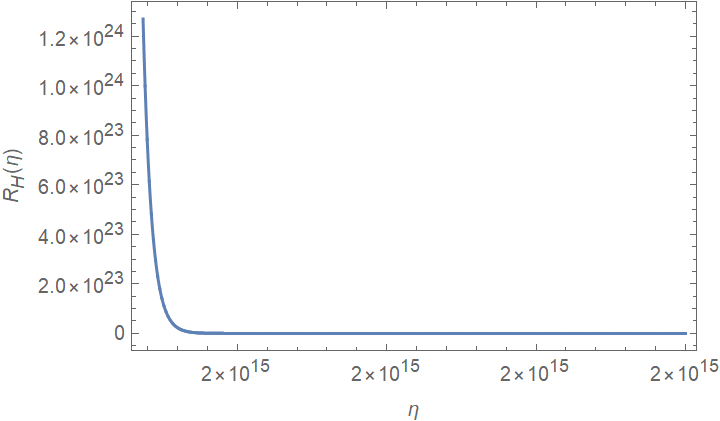}
\includegraphics[width=20pc]{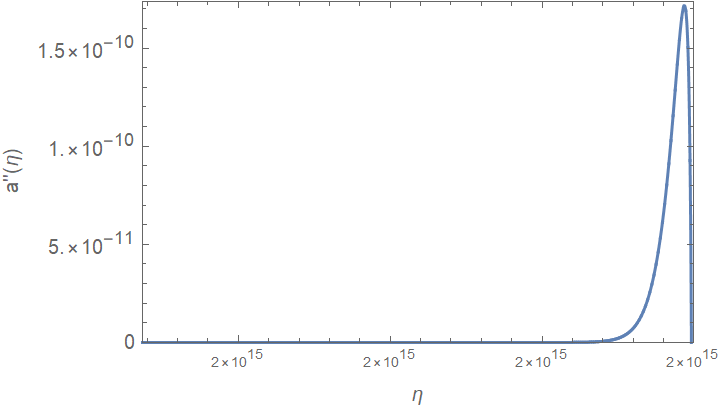}
\caption{Graphical presentation of the scale factor $a(\eta)$
(upper left), the Hubble rate (upper right) as a function of the
rescaled dimensionless time variable $\eta$, before the turnaround
point. Also in the bottom left plot we present the Hubble radius
$R_H(\eta)$ and in the bottom right the acceleration $a''(\eta)$
for the time interval for which inflation occurs
$(\eta_i,\eta_f)$. We used the values of the free parameters that
yield a viable inflationary evolution $(m,N)=(6,55.5)$ and
$(\beta,\eta,\lambda)=(10^{-20},10^{-3},10^{-23})$.}\label{plot2}
\end{figure}
So we shall prove that the evolution is indeed inflationary up to
$\eta_f$ and also we will prove that the Hubble horizon shrinks
during the inflationary regime, and also we shall show that the
Hubble rate changes sign at the turnaround point, which indicates
a turnaround cosmology. Recall that $m$ was assumed to be an even
number, thus after the turnaround point, the Hubble rate is,
\begin{equation}\label{hubbleasfunctioetaafter}
H(\eta)=-2^{-\frac{1}{m+1}} (m+1)^{\frac{1}{m+1}} \Big(-2 \lambda
+\beta \eta  \tau ^m \Big)^{\frac{1}{m+1}}\, .
\end{equation}
In order to make some graphical presentations of the evolution,
let us find a viable set of parameters $\beta$, $\tau$ and
$\lambda$ which make the amplitude of the scalar perturbations
(\ref{amplitudefinal}) compatible with the Planck constraint
$A_s\sim 2.19\times 10^{-9}$ \cite{Planck:2018jri}. We choose
$(m,N)=(6,55.5)$ which yielded a viable inflationary
phenomenology. A viable set of $(\beta,\eta,\lambda)$ values that
makes the model to have an amplitude of scalar perturbations
compatible with the Planck constraint is
$(\beta,\eta,\lambda)=(10^{-20},10^{-3},10^{-23})$, and we chose
$\lambda$ so small because physically, the inflationary era occurs
at cosmic times of the order $t\sim \mathcal{O}(10^{-25})$sec. In
Fig. \ref{plot2} we plot the behavior of the scale factor before
the turning point and after, the behavior of the Hubble rate
before and after the turning point, the evolution of the Hubble
radius $R_H=\frac{1}{H(\eta) a(\eta)}$ and the acceleration
$a''(\eta)$ in the time interval $(\eta_i,\eta_f)$. As it can be
verified from Fig. \ref{plot2}, the evolution is indeed
inflationary in the time interval $\Delta \eta=(\eta_i,\eta_f)$.
Also notice the plots that the values $\eta_i$ and $\eta_f$ are
very close to the turnaround point $\eta_T$\footnote{In this
rescaled system, due to small numbers, the difference is very
small, in fact the $\eta_f$ is $\eta_f=1.999999999999943698\times
10^{15}$ and $\eta_i=1.81544043137174317001948615\times 10^{15}$,
so these are very small differences in this rescaled time frame,
this is why the $x$-axis in the plots shows $2\times 10^{15}$}. If
we add $\kappa$ in natural units, then the order of magnitude of
$t_i$ would be $\mathcal{O}(10^{-25})$sec approximately. But this
is just a graphical presentation, the physical picture is clear.

Now viewed as dynamical system, the analytic scalar field
inflation model we developed is quite simple. The phase space is
reduced to simple curves differing only in the initial conditions
used each time. The $\dot{H}-H$ system closes under the equation,
\begin{equation}
\dot{H} = -\frac{\kappa^2}{2} \gamma H^{-m}.
\end{equation}
This equation is autonomous in $H$ alone and it defines a
one-dimensional invariant manifold, determined by,
\begin{equation}
\mathcal{M}:\qquad \dot{\phi}^2 - \gamma H^{-m} = 0.
\end{equation}
The flow never leaves this surface once initial data lie on it,
thus the system of the above equations, reduces the initial
three-dimensional phase space $(\phi,\dot{\phi},H)$ of a minimally
coupled scalar field theory, to an effectively one-dimensional
flow. Let us analyze the trajectories in the $\phi-\dot{\phi}$
plane, for various values of $H_0$, $\beta$, $m$ and $\phi_0$.
Note that in ordinary scalar field theory, the phase space
analysis is impossible to do for general potentials, and it is
done only for exponential potentials. Thus the general
trajectories for general scalar field theories in the $H-\dot{H}$
and in the $\phi-\dot{\phi}$ planes cannot be found. In the case
at hand, this is a trivial plot, since the trajectories are unique
curves, differing only on the initial conditions, and the
parameter values. To this end we will rescale the scalar field
$\phi=\frac{\varphi}{\kappa}$ where $\varphi$ is a dimensionless
variable.
\begin{figure}
\centering
\includegraphics[width=25pc]{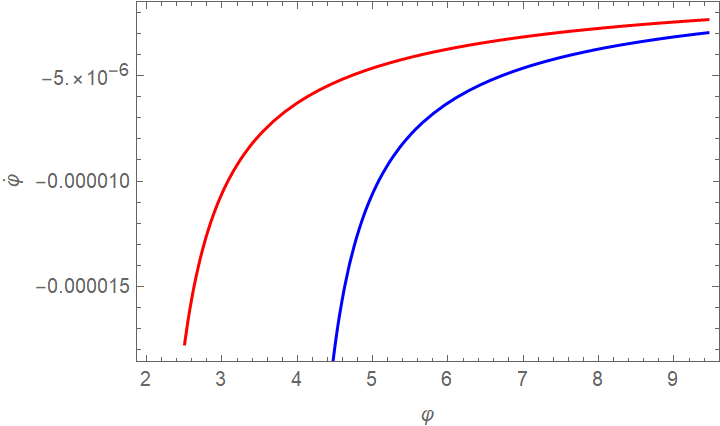}
\caption{The trajectories in the $\dot{\varphi}-\varphi$ space,
for $\varphi=2$ (red curve) and $\varphi=4$ (blue curve), in the
range $(\varphi_i,\varphi_f)$. We used the values of the free
parameters that yield a viable inflationary evolution
$(m,N)=(6,55.5)$ and
$(\beta,\eta,\lambda)=(10^{-20},10^{-3},10^{-23})$.}\label{plot3}
\end{figure}
We can easily find by eliminating the cosmic time between $\phi$
and $\dot{\phi}$, that the derivative of the scalar field as a
function of the scalar field is given as,
\begin{equation}\label{scalarfieldequation}
\dot{\varphi}=-\frac{4^{\frac{m}{m+2}}\, (m+2)^{-\frac{m}{m+2}}
\beta ^{\frac{1}{m+2}} \,\tau ^{\frac{m}{m+2}}\, \Big(\varphi
-\varphi_0\Big)^{-\frac{m}{m+2}}}{\kappa ^2}\, .
\end{equation}
For $m=6$ and for the values of the free parameters we used
earlier, we can easily find the initial and final values of the
scalar field, which are,
\begin{equation}\label{phii}
\phi_i=\frac{7.44988}{\kappa }+\phi_0\, ,
\end{equation}
\begin{equation}\label{phif}
\phi_f=\frac{0.0197297}{\kappa }+\phi_0\, ,
\end{equation}
where $\phi_0$ is the initial condition of the scalar field. In
Fig. \ref{plot3} we plot the trajectories in the
$\dot{\varphi}-\varphi$ space, for $\varphi=2$ (red curve) and
$\varphi=4$ (blue curve), in the range $(\varphi_i,\varphi_f)$. As
it can seen, the trajectory is a single curve depending on the
initial condition. Also notice that $\dot{\varphi}$ is singular at
$\varphi=\varphi_0$.

\subsection{Other Analytic Solutions and Their Inflationary
Features}

At this point, it is worth discussing other analytic solutions of
Eq. (\ref{mainequation}) for various values of $m$. We shall
consider the cases $m=0$ and $m=-2$ separately from the previous
formalism. Let us consider first the equation,
\begin{equation}\label{extra1}
\dot{\phi}^2=\gamma\, ,
\end{equation}
so from the Raychaudhuri equation we get,
\begin{equation}\label{extra2}
2\dot{H}=-\kappa^2 \gamma\, ,
\end{equation}
which yields an exact quasi-de Sitter evolution,
\begin{equation}\label{extra3}
H(t)=H_0-\frac{1}{2} \gamma  \kappa ^2 t\, ,
\end{equation}
and the scalar field as a function of the cosmic time is,
\begin{equation}\label{scalarfield}
\phi(t)=\phi_0-\sqrt{\gamma } t\, .
\end{equation}
The scalar field potential as a function of the scalar field is
easily found,
\begin{equation}\label{scalarpotential}
V(\phi)=-\frac{\gamma }{2}+\frac{3 \Big(\sqrt{\gamma } \kappa ^2
(\phi -\phi_0)+2 \phi_0\Big){}^2}{4 \kappa ^2}\, .
\end{equation}
Accordingly, we can find $\epsilon_1$, and $\epsilon_2=0$ and the
resulting observational indices are at leading order,
\begin{equation}\label{ns}
n_{\mathcal{S}}\simeq 1-\frac{2}{N}\, ,
\end{equation}
\begin{equation}\label{rn}
r=\frac{8}{N}-\frac{4}{N^2}\, .
\end{equation}
The resulting phenomenology is non-viable however, since although
the spectral index is accommodated easily in the Planck data, the
tensor-to-scalar ratio is very large for $N\sim 50-60$.

Let us also consider the scenario $\dot{\phi}^2=\gamma H^2$, which
is well known in the literature, since it yields an exponential
scalar field potential and a power-law inflationary evolution,
with eternal inflation. Indeed,
\begin{equation}\label{extra11}
\dot{\phi}^2=\gamma H(t)^2\, ,
\end{equation}
so from the Raychaudhuri equation we get,
\begin{equation}\label{extra21}
2\dot{H}=-\kappa^2 \gamma H(t)^2\, ,
\end{equation}
which yields an exact power-law inflationary evolution,
\begin{equation}\label{extra31}
H(t)=\frac{2}{\gamma  \kappa ^2 t-2 c_1}\, ,
\end{equation}
and the scalar field as a function of the cosmic time is,
\begin{equation}\label{scalarfield1}
\phi(t)=2 c_2-\frac{2 \log \left(\gamma  \kappa ^2 t-2
c_1\right)}{\sqrt{\gamma } \kappa ^2}\, .
\end{equation}
The scalar field potential as a function of the scalar field is
easily found,
\begin{equation}\label{scalarpotential1}
V(\phi)=-\frac{2 \left(\gamma  \kappa ^2-6\right) e^{\sqrt{\gamma
} \kappa ^2 (\varphi -2 c_2)}}{\kappa ^2}\, .
\end{equation}
This model leads to a constant EoS evolution, and it leads to an
eternal inflation behavior, which cannot be viable with the Planck
data. So we will not further discuss this case. See however, how
such an exponential inflation scenario can be combined with other
physical scenarios and solve the trans-Planckian issues of scalar
field inflation
\cite{Brandenberger:2004kx,Martin:2003kp,Martin:2000xs,Odintsov:2026prn,Odintsov:2025mqq}.

\section{Inflation Ending at a Pressure Singularity, Quantum Phenomena, the Reheating, Gravitational Waves and Primordial Black Holes}

Now let us focus on the interesting part of the singular analytic
slow-roll scenario. What we have as a physical picture is a
Universe starting from a non-singular point $t=0$ with finite
scale factor, so a non-singular initial scale factor. Accordingly,
the mini, non-singular Universe starts to expand exponentially,
reaching an inflationary era which ends before a classical turning
point $t=t_c$, where the Universe develops a pressure singularity.
Classically, if the turning point is reached, a maximum size is
reached, and beyond that, the Universe would start to contract
exponentially. Now this is where the interesting physics enter,
since once the pressure singularity is approached, the quantum
effects would become dominant and the classical solution with
$\dot{\phi}^2\sim H^{-m}$ would become inapplicable. Such an
analysis was performed in \cite{Nojiri:2005sx}, using both the
conformal anomaly and the thermal effects of the singularity
horizon. Here after we shall call this mechanism the
Nojiri-Odintsov conformal anomaly effect, for simplicity.

Before quantifying the analysis it is worth recalling the
classification of finite-time singularities, and show why the
finite-time singularity of the analytic singular inflation is a
pressure singularity. We will use the Nojiri-Odintsov-Tsujikawa
classification of finite-time singularities
Ref.~\cite{Nojiri:2005sx}, so the cosmic time singularities are
classified as follows (see also the review \cite{deHaro:2023lbq}),
\begin{itemize}
\item Type I (``Big Rip'')
Ref.~\cite{Caldwell:2003vq,Nojiri:2003vn,Elizalde:2004mq,Faraoni:2001tq,
Singh:2003vx,
Wu:2004ex,Sami:2003xv,Stefancic:2003rc,Chimento:2003qy,
Zhang:2005eg,Dabrowski:2006dd,Nojiri:2009pf,Capozziello:2009hc}:
Typical crushing type singularity with geodesics incompleteness.
As the singularity point at $t \to t_s$ is reached, the scale
factor $a(t)$, the effective pressure $p_\mathrm{eff}$ and the
effective energy density $\rho_\mathrm{eff}$ diverge. The Universe
cannot pass smoothly through this singularity. It is a crushing
type spacelike singularity, at the spacelike hypersurface $t=t_s$
all the physical quantities that can be defined on it strongly
diverge. \item Type II (``sudden'' or ``pressure singularity''):
Milder than the Big Rip scenario, also known as a pressure
singularity
\cite{Barrow:2004xh,Nojiri:2004ip,Barrow:2004he,FernandezJambrina:2004yy,
BouhmadiLopez:2006fu,Barrow:2009df,BouhmadiLopez:2009jk,Barrow:2011ub,
Bouhmadi-Lopez:2013tua,Bouhmadi-Lopez:2013nma,
Chimento:2015gum,Cataldo:2017nck}, see also
\cite{Balcerzak:2012ae,Marosek:2018huv}. The Universe can pass
smoothly through this singularity. It is not a crushing type
spacelike singularity, and at the spacelike hypersurface $t=t_s$
the physical quantities that can be defined on it are smooth, that
is the scale factor and the energy density, but only the effective
pressure diverges at $t=t_s$, that is $a \to a_s$,
$\rho_\mathrm{eff} \to \rho_s$ and $\left|p_\mathrm{eff}\right|
\to \infty$. \item Type III : Stronger than the Type II, the
effective pressure and the effective energy density diverge and
the scale factor remains finite. \item Type IV : The smoothest of
all the finite-time singularities
\cite{Nojiri:2004pf,Nojiri:2005sx,Nojiri:2005sr,Barrow:2015ora,Nojiri:2015fra,
Nojiri:2015fia,Odintsov:2015zza,Oikonomou:2015qha}. In this case,
all the physical quantities remain finite on the spacelike
hypersurface $t = t_s$. For this singularity, the graceful exit
from the inflationary era may be occur due to this soft
singularity \cite{Odintsov:2015gba}.
\end{itemize}
Regarding the total pressure and total energy density
$\rho_\mathrm{eff}$ and $p_\mathrm{eff}$, these are defined as,
\begin{equation}
\label{IV} \rho_\mathrm{eff} \equiv \frac{3}{\kappa^2} H^2 \, ,
\quad p_\mathrm{eff} \equiv - \frac{1}{\kappa^2} \left( 2\dot H +
3 H^2 \right)\, ,
\end{equation}
which may receive contributions in our case from the scalar field
and from matter fluids. During inflation the matter fluids
generate a subdominant effect and thus do not affect the evolution
$H(t)$, which is solely determined by the scalar field.

Now for the scenario of analytic singular inflation, the
singularity at the turning point is a typical pressure
singularity, since the scale factor acquires its maximum value and
is finite, the Hubble rate is zero, and thus finite, but $\dot{H}$
diverges. Note that this reminds a bounce, but it is not, since in
a bounce, we would have,
\[
H<0 \rightarrow H=0 \rightarrow H>0
\]
meaning that,
\[
\begin{array}{c|c}
\text{Phase} & \text{Evolution} \\ \hline
H<0 & \text{contraction} \\
H=0 & \text{bounce} \\
H>0 & \text{expansion}
\end{array}
\]
In our case the sequence is,
\[
H>0 \rightarrow H=0 \rightarrow H<0 ,
\]
which is a cosmological turnaround behavior. Now it is important
to note that in the context of the scenario we proposed for
inflation, the EoS behaves as
\begin{equation}\label{eossingular}
w_{eff}=-1+\frac{1}{3} \gamma  \kappa ^2 2^n (m+1)^{-n}
\Big(-\gamma \kappa ^2 t+2 \omega \Big){}^{-n}\, ,
\end{equation}
thus it has a singularity at $t_*=\frac{2\omega}{\gamma \kappa
^2}$. So the model exhibits a sort of sudden singularity at the
turnaround point $t_*=\frac{2\omega}{\gamma \kappa ^2}$, which
either means that the scalar field theory is ineffective beyond
the turnaround point, or simply the Universe exhibits a sudden
singularity at that point. But apart from that, if we look at the
potential at the Appendix, it is obvious that the potential also
exhibits a singularity at $t=t_*$, so apparently, the theory has a
pole at the turnaround point. This can also be seen by looking at
Eq. (\ref{scalarfieldpotentialasfunctionofphi}) where the scalar
field potential is expressed as a function of the scalar field,
but it is less transparent compared to the $V(t)$ expression
appearing at the Appendix.

Now here is the important part from a physical point of view.
Matter fields are present during the singular analytic
inflationary regime, but are disregarded as they cause subdominant
effects and do not alter the Hubble rate. So classically, the
Universe would reach the turnaround point and would go smoothly
through the pressure singularity. However, once the pressure
singularity is approached, the curvature invariants that contain
$\dot{H}$ strongly diverge. Thus the quantum effects should be
taken into account, or even thermal effects as in Ref.
\cite{Nojiri:2020sti}. Taking into account the thermal effects
would eliminate the classical singular behavior and this is a well
known feature in the literature
\cite{Nojiri:2004ip,Nojiri:2004pf,Kamenshchik:2013naa,Bates:2010nv,Tretyakov:2005en,
Calderon:2004bi,Carlson:2016iuw,Nojiri:2010wj}. Thus the classical
trajectory $\dot{\phi}^2\sim H^{-m}$ would not longer determine
the evolution of the Universe. In this work we shall use the
Nojiri-Odintsov conformal anomaly effects \cite{Nojiri:2020sti}
and discuss how these smooth out the pressure singularity and more
importantly they generate a natural reheating mechanism for the
Universe without requiring field oscillations and numerous
couplings of the inflaton to the Standard Model particles.

The mechanism is simple, when quantum matter fields are included,
the gravitational dynamics is described by the semiclassical
Einstein equations,
\begin{equation}
G_{\mu\nu} = 8\pi G \left( T_{\mu\nu}^{\mathrm{class}} + \langle
T_{\mu\nu}^{\mathrm{quantum}} \rangle \right).
\end{equation}
The quantum expectation value of the stress tensor can be
decomposed schematically as follows,
\begin{equation}
\langle T_{\mu\nu}^{\mathrm{quantum}} \rangle =
T_{\mu\nu}^{\mathrm{vac}} + T_{\mu\nu}^{\mathrm{anom}} +
T_{\mu\nu}^{\mathrm{state}},
\end{equation}
with the contributions corresponding to, vacuum polarization, the
conformal anomaly, and state-dependent particle production. In
many cosmological applications only the anomaly term is retained
because it is universal and independent of the quantum state and
this is what we adopt here too. Classically, conformally invariant
fields, that is, classically massless fields, satisfy,
\begin{equation}
T^\mu_{\ \mu} = 0.
\end{equation}
Quantum mechanically this relation is violated due to
renormalization effects, yielding the so-called trace anomaly,
\begin{equation}
\langle {T^\mu_{\mu}}_A \rangle = b F + b' \mathcal{G} + b'' \Box
R ,
\end{equation}
with, $F$ being the square of the Weyl tensor, $\mathcal{G}$ being
the Gauss-Bonnet invariant, and $R$ being the Ricci scalar.
Examples of conformal  fields that contribute to the conformal
anomaly include,  photons,  gluons,  massless fermions, and
perhaps extra conformally coupled scalar fields (string theory
remnants). Let us further elaborate on the calculation of the
conformal anomaly for the case at hand, and how the
Nojiri-Odintsov conformal anomaly smooths the pressure singularity
in the original scenario we had. The conformal anomaly $T_A$ is,
\begin{equation}
\label{OVII} T_A=b\left(\mathcal{F} + \frac{2}{3}\Box R\right) +
b' \mathcal{G} + b''\Box R\, ,
\end{equation}
with $\mathcal{F}$ being the square of the 4D Weyl tensor, and
also $\mathcal{G}$ is the Gauss-Bonnet invariant, which are given
by
\begin{equation}
\label{GF} \mathcal{F} = \frac{1}{3}R^2 -2 R_{\mu\nu}R^{\mu\nu} +
R_{\mu\nu\rho\sigma}R^{\mu\nu\rho\sigma}\, , \quad \mathcal{G}=R^2
-4 R_{\mu\nu}R^{\mu\nu} +
R_{\mu\nu\rho\sigma}R^{\mu\nu\rho\sigma}\, .
\end{equation}
Let the conformal matter fluids consist of $N$ scalars, $N_{1/2}$
spinors, $N_1$ vector fields, $N_2$ ($=0$ or $1$) gravitons, and
also $N_\mathrm{HD}$ higher-derivative conformal scalars. Then the
factors $b$ and $b'$ are equal to,
\begin{equation}
\label{bs} b= \frac{N +6N_{1/2}+12N_1 + 611 N_2 -
8N_\mathrm{HD}}{120(4\pi)^2} \, ,\quad b'=-
\frac{N+11N_{1/2}+62N_1 + 1411 N_2 -28
N_\mathrm{HD}}{360(4\pi)^2}\ ,
\end{equation}
and thus $b$ is positive and $b'$ is negative for the usual
perfect fluid conformal matter fields. Now schematically, away
from the singularity we have,
\begin{equation}
T_A \sim \frac{R^2}{(4\pi)^2},
\end{equation}
but near the singularity we have,
\begin{equation}
T_A \sim H^{(3)}, \quad H \ddot H, \quad \dot H^2 \, ,
\end{equation}
hence, if,
\begin{equation}
H^{(3)} \gg H^2 ,
\end{equation}
the anomaly contribution can dominate the dynamics of the
Universe, even if classical matter densities are negligible. The
semiclassical trace equation has the following form,
\begin{equation}
-R = 8\pi G (T_{\phi} + T_A)\, ,
\end{equation}
and near a Type II singularity one typically finds
\begin{equation}
T_A \gg R,
\end{equation}
since,
\begin{equation}
\Box R \sim (t_s - t)^{-\beta - 2}\, ,
\end{equation}
which diverges faster than
\begin{equation}
R \sim (t_s - t)^{-\beta}\, ,
\end{equation}
where we defined $\beta$ to be,
\begin{equation}\label{betadef}
\beta=-\frac{1}{m+1}
\end{equation}
from Eq. (\ref{mainsolution}). Consequently the classical singular
solution cannot satisfy the semiclassical equations, once the
anomaly effect becomes dominant. Thus the quantum backreaction
modifies the evolution and may remove the singularity. This is
basically the Nojiri-Odintsov conformal anomaly mechanism. Let us
quantify this mechanism even further, so,
\begin{equation}
\label{CA1} R = - \frac{\kappa^2}{2} \left(T_\mathrm{matter} + T_A
\right)\, .
\end{equation}
For a FRW Universe, $\mathcal{F}$ and $\mathcal{G}$ are,
\begin{equation}
\label{CA2} \mathcal{F}=0\ ,\quad \mathcal{G}=24\left(\dot H H^2 +
H^4\right)\ .
\end{equation}
We assumed that $H$ behaves  in the following way,
\begin{equation}
\label{R13} H \sim \left(t_s - t\right)^{-\beta}\, ,
\end{equation}
and recall $\beta=-\frac{1}{m+1}$ from Eq. (\ref{betadef}). Since,
$0> \beta > -1$ the singularity is a  Type II or pressure
singularity. In this case, we find that $\mathcal{G}\sim 24 \dot H
H^2 \sim \left(t_s - t \right)^{-3\beta -1}$. Since $R \sim
\left(t_s - t \right)^{-\beta - 1}$, the Gauss-Bonnet term in
$T_A$ is less singular compared with $R$. However, as we mentioned
earlier, $\Box R$ behaves as $\Box R \sim \left(t_s -
t\right)^{-\beta - 3}$, thus it is more singular than the scalar
curvature. Hence if $2b/3 + b''\neq 0$, the contribution from
$T_A$ in the trace equation (\ref{CA1}) is more singular that $R$
near the pressure singularity. This simple dominant balance
analysis, simply shows that the solution (\ref{mainsolution}) no
longer satisfies the trace equation (\ref{CA1}), thus the Type II
singularity is avoided due to the quantum effects near the
pressure singularity.

Now this pressure singularity actually offers an elegant reheating
mechanism, without resorting to scalar field oscillations and
having the awkward necessity of coupling the inflaton differently
to all the Standard Model particles. The line of thinking is the
following, quantum particle production occurs when the background
spacetime evolves rapidly. In an expanding Universe, the
production rate of particles roughly scales with the effective
frequency variation of modes, which is governed by the scale
factor. Dimensionally one finds,
\begin{equation}
\Gamma \sim \left| \frac{a''}{a} \right|,
\end{equation}
where primes denote derivatives with respect to conformal time.
Equivalently, in terms of the cosmic time, this corresponds
approximately to,
\begin{equation}
\Gamma \sim |\dot H| + H^2 .
\end{equation}
Near a sudden (Type II) singularity the second derivative of the
scale factor diverges,
\begin{equation}
\ddot a \rightarrow \infty ,
\end{equation}
which implies,
\begin{equation}
\frac{a''}{a} \rightarrow \infty .
\end{equation}
Consequently, the vacuum modes are excited violently and particle
production becomes extremely efficient. The produced particles
generate radiation and can significantly affect the cosmic
evolution. The resulting cosmological scenario can be summarized
as the following sequence, no-singular emergence of the Universe,
followed by a slow-roll inflationary regime, after which the
Universe approaches a sudden Singularity, then quantum effects
smooth out the singularity, causing severe particle production,
which generates a radiation domination era. The Universe after
that evolves classically. The energy density of produced particles
can be estimated as,
\begin{equation}
\rho_r \sim \int |\beta_k|^2 \, k \, d^3k ,
\end{equation}
where \(\beta_k\) are Bogoliubov coefficients describing particle
production. When derivatives of the curvature become large,
\(|\beta_k|^2\) increases significantly and the radiation energy
density grows rapidly, producing backreaction on the expansion,
thus we have,
\begin{equation}
\rho_r \sim T^4 ,
\end{equation}
which effectively reheats the Universe.

However, there are more important consequences in the Universe due
to the pressure singularity. Specifically, near the pressure
singularity, the scalar perturbations would be enhanced and thus
primordial black holes may be formed and in addition secondary
gravitational waves may be generated with observable features in
the energy spectrum from future gravitational wave experiments.
Let us qualitatively analyze these in brief.  Specifically, the
scalar perturbations obey the Mukhanov-Sasaki equation,
\begin{equation}
v_k'' + \left(k^2 - \frac{z''}{z}\right) v_k = 0 ,
\end{equation}
and in our case,
\begin{equation}
z = a \frac{\dot\phi}{H}.
\end{equation}
Near a Type II singularity derivatives of the Hubble parameter
diverge, which implies
\begin{equation}
\frac{z''}{z} \sim (t_s - t)^{-2}.
\end{equation}
This produces a sharp spike in the power spectrum of the
perturbations, leading to strong amplification of modes and
possibly enhanced particle production. If the amplification of
curvature perturbations becomes sufficiently large,
\begin{equation}
P_\zeta(k) \sim 10^{-2},
\end{equation}
the resulting density fluctuations can collapse upon horizon
re-entry and form primordial black holes. Apparently we are
considering very small wavelength modes, immediately entering the
horizon after the inflationary regime, and before the singularity
is reached, so from $t=t_f$ to $t=t_{c}$. Thus a sudden
singularity after the end of inflation may generate primordial
black holes, on very small scales without affecting CMB
observables. Such features typically occur on scales much smaller
than those probed by the CMB. Therefore the relevant observational
probes include  primordial black holes,  induced gravitational
waves. Regarding the secondary gravitational waves, there are
fundamental works appearing in the literature
\cite{Sasaki:2025zao,Sasaki:2025vql,Gorji:2023ziy,Ota:2022hvh,
Domenech:2021wkk,Zhou:2020kkf,Domenech:2020kqm,Pi:2020otn,
Kuroyanagi:2017kfx,Wu:2021zta,Gao:2021vxb,Gao:2020tsa,
Ragavendra:2020sop,Lin:2020goi,Lu:2019sti}, see also
\cite{Zhang:2018dvc,Ye:2022tgs} for regularization techniques, and
the logic in our context is the following, the evolution equation
for tensor modes $h_{ij}$ is,
\begin{equation}
h_{ij}'' + 2\mathcal{H} h_{ij}' - \nabla^2 h_{ij} = 4
\mathcal{T}_{ij}^{\ \ lm} S_{lm},
\end{equation}
where $\mathcal{H}=a'/a$ is the conformal Hubble parameter and
$\mathcal{T}_{ij}^{\ \ lm}$ is the transverse-traceless projection
operator and the source term $S_{lm}$ is quadratic in scalar
perturbations. Consequently, the enhanced scalar modes generated
near the sudden singularity would produce a stochastic background
of secondary gravitational waves. The corresponding energy density
spectrum would be approximately,
\begin{equation}
\Omega_{\rm GW}(k) \sim \mathcal{O}(1)\, P_\zeta^2(k).
\end{equation}
Therefore a burst of scalar amplification associated with a sudden
singularity could potentially lead to a peaked spectrum of
primordial gravitational waves on very small scales. Such signals
may be connected to primordial black hole formation and could be
probed by future high-frequency gravitational wave detectors.

\subsection{Inflationary Effective Field Theory Scenarios and Beyond the UV-cutoff Physics}

Another interesting interpretation near the pressure singularity
is the effective field theory of inflation
\cite{Cheung:2007st,Weinberg:2008hq,Senatore:2010wk,Kallosh:2013hoa,Galante:2014ifa}.
Specifically, we can imagine that the singular analytic scalar
field inflation scenario is a part of an effective theory of
inflation, which provides a framework for describing the dynamics
of the inflationary phase, without specifying the detailed
microphysical origin of the inflaton sector. The main idea is
that, at energies well below a physically imposed cutoff scale
$\Lambda$, the relevant degrees of freedom can be captured by an
effective action organized as an expansion in operators consistent
with the symmetries of the cosmological background. In
single-field inflation, the time evolution of the background
scalar field $\phi(t)$ spontaneously breaks time diffeomorphism
invariance, thus leaving spatial diffeomorphisms unbroken. This
allows the construction of the most general action governing the
fluctuations around the quasi-de Sitter background. At the
background level the dynamics is usually described by a canonical
scalar field minimally coupled to gravity,
\[
S=\int d^4x\sqrt{-g}\left[\frac{M_{\rm
Pl}^2}{2}R-\frac12(\partial\phi)^2-V(\phi)\right],
\]
leading to the Friedmann equations.
\[
3M_{\rm Pl}^2H^2=\frac12\dot{\phi}^2+V(\phi), \qquad \dot
H=-\frac{\dot{\phi}^2}{2M_{\rm Pl}^2},
\]
and inflation occurs when the slow-roll parameters,
\[
\epsilon=-\frac{\dot H}{H^2}, \qquad
\eta=\frac{\dot\epsilon}{H\epsilon}
\]
remain small, $\epsilon\ll1$ and $|\eta|\ll1$, so that the
expansion is approximately quasi-de Sitter. Within the effective
field theory approach one constructs the most general action for
perturbations around the inflationary background by fixing a
unitary gauge, in which the scalar fluctuations are absorbed into
the metric. The resulting action can be written schematically as,
\[
S=\int d^4x\sqrt{-g}\left[\frac{M_{\rm Pl}^2}{2}R - M_{\rm
Pl}^2\dot H g^{00} - M_{\rm Pl}^2(3H^2+\dot H) +\sum_{n\ge2}
\frac{M_n^4}{n!}(\delta g^{00})^n + \cdots \right],
\]
with the higher-order operators encoding the effects of unknown
high-energy physics well above the cutoff scale. This formalism
makes explicit that inflation should be regarded as an effective
description, which is valid only within a specific evolutionary
regime. The effective field theory perspective is particularly
useful in our case, in which the scalar field derivative develops
a pressure singularity after the inflationary era. In our case,
after inflation and near the pressure singularity, the kinetic
energy grows rapidly as the cosmic time approaches the turnaround
point, thus driving the system toward the boundary of validity of
the effective theory. This for example occurs when,
\[
\dot{\phi}^2 \sim \Lambda^4,
\]
and thus the higher-order operators in the effective action,
\[
\mathcal{L} = \frac12(\partial\phi)^2
+\frac{1}{\Lambda^4}(\partial\phi)^4
+\frac{1}{\Lambda^8}(\partial\phi)^6+\cdots,
\]
which were disregarded, become dynamically more important and the
canonical scalar description breaks down. From this viewpoint,
singular behavior in the classical background evolution can be
interpreted not as a fundamental physical singularity, but as an
indication that the additional operators or extra effective field
theory degrees of freedom must be taken into account for the
dynamics. Hence, the effective field theory of inflation provides
a robust framework for analyzing inflationary models, while at the
same time being model agnostic about the UV completion of the
theory. It may also justify how non-crushing type singularities
may signal the breakdown of the low-energy classical scalar field
description, and thus point out that the singularity is not a
pathology of the fundamental theory, but an indication that the
classical description breaks down once the kinetic energy of the
scalar field becomes comparable in magnitude with the cut-off.

\subsection{Inflationary Phase Transitions near the Pressure Singularity?}

An interesting interpretation of the pole in $\dot{H}$ and the
kinetic energy of the scalar field $\dot{\phi}^2$ appearing at the
turnaround point $t=t_c$, is that it may actual indicate that a
phase transition in the scalar sector takes place, rather than a
true physical singularity. In this view the relation
\[
\dot{\phi}^{2}=\gamma H^{-m}
\]
is valid only for a particular effective phase of the total scalar
field theory. As the cosmological evolution drives the system
toward
\[
H\rightarrow0,
\]
the kinetic energy grows rapidly,
\[
\dot{\phi}^{2}\rightarrow \infty,
\]
a fact that indicates, that the system approaches the boundary of
validity of that phase. In many effective field theories, a
divergence of this sort indicates that the theory is approaching a
critical point, where new degrees of freedom become relevant. The
scalar sector may undergo a transition in which the effective
potential, the kinetic structure, or the vacuum state of the field
changes. Quantitatively this can be simply realized by a scalar
potential of the form,
\[
V(\phi,\chi)=V_1(\phi)+g(\chi)\phi^2,
\]
with the auxiliary field or background quantity $\chi$ evolving
with cosmic time. Such a formulation is also used in standard
power-law inflationary scenarios to end the eternal inflation
phase \cite{Odintsov:2026prn,Odintsov:2025mqq}. As the Universe
expands ,and the Hubble parameter decreases, the effective mass of
the scalar field,
\[
m_{\rm eff}^2(\chi)=\frac{\partial^2 V}{\partial \phi^2},
\]
may change sign, so when,
\[
m_{\rm eff}^2=0
\]
the system reaches a critical point and the vacuum configuration
of the scalar field changes. Such a phase transition can occur
when the cosmological evolution reaches a specific value of the
Hubble parameter. In this theoretical framework, the divergence of
$\dot{\phi}^2$ near the point where $H=0$, could indicate that the
field is being driven rapidly toward the new vacuum configuration.
Instead of reaching an actual singularity, the scalar sector
undergoes a dynamical transition that drastically modifies the
effective theory. The original evolution,
\[
\dot{\phi}^{2}=\gamma H^{-m}
\]
is then replaced by a different dynamical evolution governed by
the new phase. But this scenario is just another perspective, and
one must rely on oscillations of the scalar fields in the new
phase to explain reheating. Thus one must couple again the scalar
field(s) with the Standard Model particles to explain the
reheating and generate the radiation domination era. We believe
that the Nojiri-Odintsov conformal anomaly approach for the
reheating of the Universe in our context, is much more appealing
theoretically. We aimed though to show the inflationary phase
transitions perspective, also appearing in the literature in
various forms
\cite{Zou:2026wzi,Bao:2026bgu,Hu:2025xdt,An:2023jxf,An:2022toi,Kamada:2011bt,Abolhasani:2010kn,Itzhaki:2008hs,Dimopoulos:2022mce,Gong:2022tfu,Dimopoulos:2017qqn,Kawasaki:2015ppx,Lyth:2012yp,Fonseca:2010nk}

\section{Conclusions}

In this work we presented a class of minimally coupled scalar
field models, which can realize a slow-roll era and can be solved
fully analytically. We examined the solutions of the model, and we
found that the models are essentially one parameter models and can
be compatible with the ACT data. In fact in limiting cases, the
models predict a bluer tilt of the spectral index near the value
0.98 and the tensor-to-scalar ratio tends to very small, nearly
zero values. The cosmological solution describes a Universe which
starts from a non-singular state and expands, realizing a
slow-roll era. After the inflationary era ends, the cosmological
system develops a pressure singularity at least classically, and
the Universe reaches a maximum size and starts to contract.
However, if we take into account the quantum phenomena near the
singularity, the classical singularity is avoided, and more
importantly these quantum anomalies induce a reheating mechanism
which avoids oscillating scalar fields and its numerous couplings
to the Standard Model particles. Thus the models we presented have
the attribute of being able to study analytically, describe an
initially non-singular emergent Universe, distinct from other
emergent Universe scenarios
\cite{Mulryne:2005ef,Ellis:2002we,Ellis:2003qz}, the model
produces a slow-roll inflationary era and develops a pressure
singularity classically. The quantum phenomena near the
singularity make the Universe avoid the singularity, and at the
same time they provide a natural reheating mechanism after the
inflationary era. In addition, the power spectrum might be
enhanced, and thus the generation of primordial gravitational
waves and of secondary gravitational waves is also a possibility.

Future perspectives of this work is including thermal horizon
effects near the sudden singularity and extending the theoretical
framework to k-inflationary theories, $F(R,\phi)$ gravity or even
Einstein-Gauss-Bonnet gravity. We hope to address some of these
tasks soon.

\section*{Appendix: Definition of Constants and of Lengthy Expressions}

In this appendix we quote the scalar field potential as a function
of the cosmic time. This can be found by using the Friedmann
equation (\ref{fr1}), and the solutions for the scalar field
$\phi(t)$ given in Eq. (\ref{mainsolutionscalarfield}) and the
Hubble rate (\ref{mainsolution}), so the scalar potential reads,
\begin{equation}\label{scalarpotentialastime}
V(t)=4^{-\frac{1}{m+1}} \left(\frac{3 (m+1)^{\frac{2}{m+1}} \Big(2
\omega-\gamma  \kappa ^2 t\Big)^{\frac{2}{m+1}}}{\kappa ^2}-\gamma
2^{\frac{1}{m+1}} (m+1)^{-\frac{m}{m+1}} \Big(2 \omega-\gamma
\kappa ^2 t\Big)^{-\frac{m}{m+1}}\right)\, .
\end{equation}
Also here we shall define the constants appearing in Eq.
(\ref{scalarfieldpotentialasfunctionofphi}) which are,
\begin{equation}\label{constantk1}
\mathcal{K}=2^{\frac{3 m^2+10 m+6}{2 (m+1)^2}}
\end{equation}
and
\begin{equation}\label{constantk2}
\mathcal{B}=2^{\frac{-5 m-4}{(m+1)^2}} (m+2)^{-\frac{2 m}{m+2}}
\gamma ^{\frac{2}{m+2}} \kappa ^{-\frac{4 m}{m+2}}\, .
\end{equation}
Finally, the parameter $\Gamma$ appearing in Eq. (\ref{ti}) is
defined as follows,
\begin{equation}\label{gammaparameter}
\Gamma=-\frac{2^{-\frac{1}{m+1}} (m+1)^{n} \Big(2 \omega-\gamma
\kappa ^2 t_f\Big)^{n}}{\gamma  \kappa ^2 (m+2)}\, .
\end{equation}

\end{document}